\documentclass[12pt,preprint]{aastex}

\usepackage{amsmath}

\shorttitle{MAXI J0158$-$744 as ignition of nova}
\shortauthors{Morii et al.}

\begin{document}


\title{Extraordinary luminous soft X-ray transient MAXI J0158$-$744
as an ignition of a nova on a very massive O-Ne white dwarf}


\author{M. Morii\altaffilmark{1},
H.~Tomida\altaffilmark{2},
M.~Kimura\altaffilmark{2},
F.~Suwa\altaffilmark{3},
H.~Negoro\altaffilmark{3},
M.~Serino\altaffilmark{1},
J.~A.~Kennea\altaffilmark{4},
K.~L.~Page\altaffilmark{5},
P.~A.~Curran\altaffilmark{6},
F.~M.~Walter\altaffilmark{7},
N.~P.~M.~Kuin\altaffilmark{8},
T.~Pritchard\altaffilmark{4},
S.~Nakahira\altaffilmark{2},
K.~Hiroi\altaffilmark{9},
R.~Usui\altaffilmark{10},
N.~Kawai\altaffilmark{10},
J.~P.~Osborne\altaffilmark{5},
T.~Mihara\altaffilmark{1},
M.~Sugizaki\altaffilmark{1},
N.~Gehrels\altaffilmark{11},
M.~Kohama\altaffilmark{2},
T.~Kotani\altaffilmark{12},
M.~Matsuoka\altaffilmark{1},
M.~Nakajima\altaffilmark{13},
P.~W.~A.~Roming\altaffilmark{14},
T.~Sakamoto\altaffilmark{15},
K.~Sugimori\altaffilmark{10},
Y.~Tsuboi\altaffilmark{16},
H.~Tsunemi\altaffilmark{17},
Y.~Ueda\altaffilmark{9},
S.~Ueno\altaffilmark{2},
and
A.~Yoshida\altaffilmark{15}
}

\affil{$^{1}$ MAXI team, Institute of Physical and Chemical Research (RIKEN),
  2-1 Hirosawa, Wako, Saitama 351-0198, Japan.}
\affil{$^{2}$ ISS Science Project Office, Institute of Space and Astronautical
  Science, Japan Aerospace Exploration Agency, 2-1-1 Sengen, Tsukuba,
  Ibaraki 305-8505, Japan.}
\affil{$^{3}$ Department of Physics, Nihon University, 1-8-14 Surugadai, Chiyoda,
  Tokyo 101-8308, Japan.}
\affil{$^{4}$ Department of Astronomy and Astrophysics,
  The Pennsylvania State University,
  525 Davey Laboratory, University Park, Pennsylvania 16802, USA.}
\affil{$^{5}$ Department of Physics and Astronomy, University of Leicester,
  University Road, Leicester LE1 7RH, UK.}
\affil{$^{6}$ International Centre for Radio Astronomy Research, Curtin University,
  GPO Box U1987, Perth, WA 6845, Australia.}
\affil{$^{7}$ Department of Physics and Astronomy, Stony Brook University, Stony Brook,
  NY 11794-3800, USA.}
\affil{$^{8}$ Mullard Space Science Laboratory, University College London,
  Holmbury St Mary, Dorking, Surrey RH5 6NT, UK.}
\affil{$^{9}$ Department of Astronomy, Kyoto University, Oiwake-cho, Sakyo-ku,
  Kyoto 606-8502, Japan.}
\affil{$^{10}$ Department of Physics, Tokyo Institute of Technology,
  Ookayama 2-12-1,
  Meguro-ku, Tokyo 152-8551, Japan.}
\affil{$^{11}$ NASA Goddard Space Flight Center, Greenbelt, Maryland 20771, USA.}
\affil{$^{12}$ Waseda University, Organization for University Research Initiatives,
  17 Kikuicho, Shinjuku, Tokyo 162-0044, Japan.}
\affil{$^{13}$ School of Dentistry at Matsudo, Nihon University, 2-870-1 Sakaecho-nishi,
  Matsudo, Chiba 271-8587, Japan.}
\affil{$^{14}$ Southwest Research Institute, Space Science and Engineering Division,
  PO Drawer 28510, San Antonio, Texas 78228-0510, USA.}
\affil{$^{15}$ Department of Physics and Mathematics, Aoyama Gakuin University,
  5-10-1 Fuchinobe, Chuo-ku, Sagamihara, Kanagawa 252-5258, Japan.}
\affil{$^{16}$ Department of Physics, Faculty of Science and Engineering,
  Chuo University, 1-13-27 Kasuga, Bunkyo-ku, Tokyo 112-8551, Japan.}
\affil{$^{17}$ Department of Earth and Space Science, Osaka University,
  1-1 Machikaneyama, Toyonaka, Osaka 560-0043, Japan.}

\begin{abstract}
We present the observation
of an extraordinary luminous soft X-ray transient,
MAXI J0158$-$744, by the Monitor of All-sky X-ray Image (MAXI) on 2011 November 11.
This transient is characterized by a soft X-ray spectrum,
a short duration ($1.3 \times 10^{3}$~s $< \Delta T_d < 1.10 \times 10^{4}$~s),
a very rapid rise ($< 5.5 \times 10^{3}$~s),
and a huge peak luminosity of $2 \times 10^{40}$ erg s$^{-1}$
in 0.7$-$7.0~keV band.
With {\it Swift} observations and optical spectroscopy from
the Small and Moderate Aperture Research Telescope System (SMARTS),
we confirmed that the transient is a nova explosion,
on a white dwarf in a binary with a Be star,
located near the Small Magellanic Cloud.
An extremely early turn-on of the super-soft X-ray source (SSS) phase 
($< 0.44$~d), the short SSS phase duration of about one month, and
a 0.92~keV neon emission line found in the third MAXI scan,
1296~s after the first detection, suggest that
the explosion involves a small amount of ejecta and
is produced on an unusually massive O-Ne white dwarf close to,
or possibly over, the Chandrasekhar limit.
We propose that the huge luminosity detected with MAXI was
due to the fireball phase,
a direct manifestation
of the ignition of the thermonuclear runaway process in a nova explosion.

\end{abstract}





\section{Introduction}\label{sec: intro}

Classical or recurrent novae are typically characterized
by a rapid optical increase of 6 magnitudes or more
followed by a decline to quiescence over the next $3-300$ days \citep{Warner_1995}.
They originate from an accreting binary system consisting of a white dwarf (WD)
and a mass-losing late-type companion star.
Novae are triggered by thermonuclear runaways (TNR) 
lasting $\sim 100$~s at the bottom of
the accreted mass layer on the WD surface \citep{Warner_1995,Starrfield_Iliadis_Hix_2008}.
The TNR blows off the outer layer of the accumulated mass and
causes an optically thick wind expanding up to
$\sim 100 R_\odot$. 
It produces bright blackbody emission ($\sim 10^{38}$ erg s$^{-1}$,
comparable to the Eddington luminosity
of a $1 M_\odot$ object)
at optical bands.
This optical nova phase lasts for $\sim 3 - 300$ days \citep{Warner_1995}.
At the same time a blast wave, caused by a nova explosion in a dense circumstellar medium,
sometimes produces shock-induced optically-thin hard X-ray emission lasting $\sim 10$ days,
as observed in RS Ophiuchi \citep{Sokoloski+2006} and V407 Cyg \citep{Nelson+2012},
for example.
After the wind stops, the photosphere shrinks down to the WD surface
($\sim 10^3 - 10^4$ km),
and the blackbody temperature increases to $\sim 100$ eV,
meaning the emission is in the soft X-ray energy range.
This transient phase with soft X-ray emission is called the super-soft source
phase (SSS phase) and it lasts about $\sim 100 - 1000$ days \citep{Schwarz+2011,Hachisu_Kato_2006}.
When the nuclear burning stops, the SSS phase ends.
Novae are classified into speed classes
by the decay time scale of their optical light curves \citep{Warner_2008}.
Faster novae show earlier turn-ons and shorter durations of the SSS phase.
For example, the fastest nova, U Sco,
showed a turn-on of the SSS phase at 10 days and 
had a duration of about 25 days
\citep{Schwarz+2011}.
In general, the evolution of classical/recurrent novae
has been established, except for the early phase.
At the time of the TNR, the very early and short emission (a few hours)
is predicted to appear in the UV to soft X-ray range, called ``fireball phase''
\citep{Starrfield+1998, Krautter_2008a, Krautter_2008b, Starrfield_Iliadis_Hix_2008}.
However, no such signal has yet been observed
due to the difficulty in detecting the abrupt short phenomenon
appearing in these energy range.


Monitor of All-sky X-ray Image \citep[MAXI;][]{Matsuoka+2009}
is an all-sky X-ray monitor, which is operated on 
the Japanese Experiment Module, the Exposed Facility (JEM-EF)
on the International Space Station (ISS).
MAXI carries two types of X-ray cameras: Gas Slit Camera
\citep[GSC;][]{Mihara+2011,Sugizaki+2011}
and Solid-state Slit Camera
\citep[SSC;][]{Tsunemi+2010,Tomida+2011}.
GSC and SSC have wide fields of view (FoVs) of $1.5^\circ \times 160^\circ$
and $1.5^\circ \times 90^\circ$, respectively, and
they scan almost all of the sky every $\sim 92$ min
utilizing the attitude rotation coupled with the ISS orbital motion
\citep[See Fig.~1 in ][]{Sugizaki+2011}.
GSC covers the $2 - 30$ keV band using gas proportional counters,
while SSC covers the $0.5 - 12$ keV band with the X-ray CCDs.
MAXI started its operation in orbit in 2009 August.

\begin{figure*}
  \begin{center}
    \includegraphics[width=15cm]{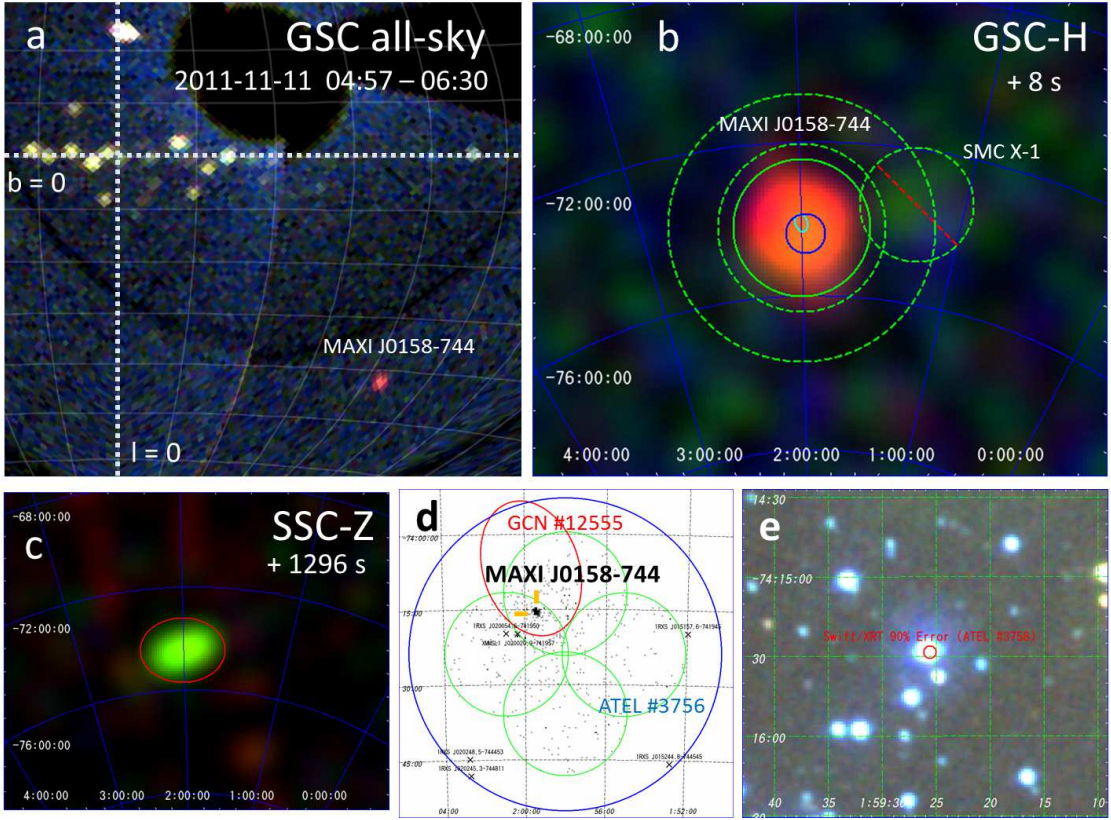}
  \end{center}
  \caption{Images of MAXI J0158$-$744 taken by MAXI and {\it Swift}.
    ({\bf a}) A part of the GSC all-sky image shown in 
    Hammer-Aitoff projection of Galactic coordinates.
    Red, green, and blue color maps represent
    the intensities in 2$-$4, 4$-$10, and 10$-$20 keV bands, respectively.
    Grid lines are drawn every 1~h and 10$^\circ$ in longitude and latitude, respectively.
    ({\bf b}) Scan image of GSC-H around $T_{\rm trig}$.
    The source and background regions for the spectral analysis are shown in
    a green solid circle and dashed annulus, respectively.
    The excluded region centered at SMC X-1 is shown by a dashed circle.
    The colors represent the same energy bands as in {\bf a}.
    The error regions reported in \citet{Kimura+2011} and \citet{Morii+2011a}
    are shown by a blue circle and cyan ellipse, respectively.
    Here, these error regions are made by adding the 90\% confidence statistical error and
    the systematic uncertainty (0.1 deg, 90\% containment radius).
    ({\bf c}) Scan image of SSC-Z around $T_{\rm trig} + 1296$~s.
    The source regions are shown by the red ellipses.
    Red, green, and blue color maps represent
    the intensities in 0.7$-$1, 1$-$4, and 4$-$10 keV bands, respectively.
    ({\bf d}) {\it Swift} XRT mosaic image around MAXI J0158$-$744 obtained by 
    the tiling observation data taken to search the X-ray counterpart \citep{Kennea+2011b}.
    Green circles show FoVs of four {\it Swift} XRT pointings.
    The X-ray sources cataloged in ROSAT All-sky Survey and XMM-Newton Slew Survey
    are marked with X symbols.
    The same error regions as {\bf b} are shown by blue circle and red ellipse.
    ({\bf e}) {\it Swift} UVOT image made by combining the data with $v$, $b$, and $u$ filters,
    colored in red, green, and blue, respectively.
    The 90\% confidence error circle obtained by {\it Swift} XRT
    is shown by a red circle.
    \label{fig: maxi_finding_chart}
  }
\end{figure*}

The MAXI transient alert system \citep{Negoro+2010}
was triggered on 2011 November 11 at 05:05:59 UT ($= T_{\rm trig}$)
by a new bright soft X-ray source near the Small Magellanic Cloud (SMC;
Fig.~\ref{fig: maxi_finding_chart}a)
\footnote{GCN Notice: http://gcn.gsfc.nasa.gov/other/39.maxi;
MAXI alert mailing list [New-transient:39]:
\hspace{0.6cm}http://maxi.riken.jp/pipermail/new-transient/2011-November/000038.html}.
We analyzed the data and reported the source position through
an Astronomer's Telegram 
\citep[ATEL;][]{Kimura+2011} and the GRB Coordinates Network
\citep[GCN;][]{Morii+2011a}.

At 0.44~days after $T_{\rm trig}$,
{\it Swift} X-ray Telescope \citep[XRT;][]{Gehrels+2004,Burrows+2005}
began follow-up observations \citep{Kennea+2011a}
with a tiling mode to cover
the MAXI error circle
\citep[Fig.~\ref{fig: maxi_finding_chart}d;][]{Kimura+2011}.
An uncataloged X-ray source was found 
within the MAXI GSC error ellipse
\citep[Fig.~\ref{fig: maxi_finding_chart}d;][]{Kennea+2011b,Morii+2011a}.
Within the {\it Swift} XRT error circle,
a single optical source is cataloged in USNO-A2.0,
which was also reported as a source
with a near-infrared excess \citep[ID 115 in][]{Nishiyama+2007}.
The position is consistent with that of an optical counterpart observed
by {\it Swift} Ultraviolet/Optical Telescope
\citep[UVOT;][]{roming2005:SSRv120},
$\alpha = 01\,^{\rm h}\,59\,^{\rm m}\,25.83\,^{\rm s}$,
$\delta = -74\,^\circ\,15\,^\prime\,27.9\,^{\prime\prime}$,
with an estimated uncertainty of 0.5 arcsec
\citep[90\% confidence; Fig.~\ref{fig: maxi_finding_chart}e;][]{Kennea+2011b}.

The {\it Swift} XRT spectra obtained after $T_{\rm trig} + 0.44$~d
were reported to be 
similar to the SSS phase of novae \citep{Li+2012}.
Further follow-up observations by {\it Swift}
and ground-based optical observations confirmed
that this source is a binary system consisting of a WD and a Be star
at the distance of the SMC \citep[$d = 60$ kpc;][]{Hilditch_Howarth_Harries_2005,Li+2012}.
\citet{Li+2012} analyzed the spectrum of the GSC scan at $+8$~s,
using the on-demand data products provided by the MAXI team,
and reported that the luminosity was $\sim 1.6 \times 10^{39}$ erg s$^{-1}$ 
in the $2-4$ keV band;
this is one order of magnitude brighter than the Eddington luminosity of a solar mass object.
To explain the huge outburst luminosity,
\citet{Li+2012} proposed a model of the interaction of the ejected nova shell
with the Be star wind in which the WD is embedded.

Here we present observations of MAXI J0158$-$744
by MAXI, {\it Swift} and 
the Small and Moderate Aperture Research Telescope System
(SMARTS)
in Section~\ref{sec: observation}.
The analysis and results of the MAXI GSC and SSC data are described in
Sections~\ref{sec: MAXI/GSC} and \ref{sec: MAXI/SSC}
with the detailed spectral analysis for the third scan of MAXI 
shown in
Sections \ref{sec: MAXI/SSC spec M+2} and \ref{sec: MAXI/SSC spec M+2 ph}.
The upper flux limits before and after the MAXI detection
are given in Section~\ref{subsec: upper limit GSC}
while the analysis and results for the {\it Swift} and SMARTS follow-up
observations are presented in Section \ref{sec: follow-up}.
The historical observations of this source
are described in Section~\ref{sec: history}.
In Section~\ref{sec: discussion},
we interpret the results obtained by MAXI, {\it Swift} and SMARTS and
discuss the emission mechanism of the very luminous soft X-ray 
transient detected by MAXI.
Finally, we summarize this paper in Section~\ref{sec: summary}.

\section{Observation}\label{sec: observation}

MAXI J0158$-$744 \citep{Kimura+2011} was
first detected during a MAXI GSC scan
(Fig.~\ref{fig: maxi_finding_chart}b),
centered at $T_{\rm trig} + 8$~s
within the $55$~s triangular transit response \citep[See Fig.~9 in ][]{Sugizaki+2011}.
It was subsequently detected twice by 
the MAXI SSC in scans at $+220$~s and $+1296$~s
(Fig.~\ref{fig: maxi_finding_chart}c).
Hereafter, we designate the MAXI scans by the mid-time of the scan transit,
referred to $T_{\rm trig}$.
Subsequent GSC scans to date (up to 2013 July 8)
have failed to detect the source.
In addition, the source had not been detected in prior GSC scans,
since MAXI observations started on 2009 August 14 up to
the previous scan at $-5530$~s.
The MAXI observations around $T_{\rm trig}$ are summarized in Table~\ref{table: maxi obs}.

\begin{table*}
  \begin{center}
    \caption{Summary of MAXI observation\label{table: maxi obs}}
    \begin{tabular}{lccccc}
      \tableline \tableline
      Scan-ID & Scan Time(Start -- End)(UT) & $\Delta t$(s)$^a$ & $t$(s)$^b$ & Detector & Flux$^{c}$ \\ 
      \tableline
      M$-1$$^d$ & 2011-11-11 03:33:22 -- 03:34:17$^{\ast}$ & 55 & $-5530$  & GSC-H &
      $< 2.75 \times 10^{-9}$$\,\,\,\,^{e,f}$ \\
      M+0       & 2011-11-11 05:05:39$^{\dagger}$ -- 05:06:34 & 55 & $+8$  & GSC-H &
      $1.81^{+0.14}_{-0.33} \times 10^{-8}$$\,\,\,\,^{g}$ \\
      M+1       & 2011-11-11 05:09:13 -- 05:10:04 & 51 & $+220$  & SSC-H &
      $4.16^{+0.47}_{-1.20} \times 10^{-8}$$\,\,\,\,^{g}$ \\
      M+2       & 2011-11-11 05:27:09 -- 05:28:00$^{\ddagger}$ & 51 & $+1296$ & SSC-Z & 
      $1.57^{+0.17}_{-0.18} \times 10^{-8}$$\,\,\,\,^{h}$ \\
      M+3       & 2011-11-11 06:37:56$^{\S}$ -- 06:38:51 & 55 & $+5545$ & GSC-H &
      $< 9.60 \times 10^{-10}$$\,\,\,\,^{e,f}$ \\
      \tableline
    \end{tabular}
    \tablenotetext{a}{Total length of the MAXI scan.}
    \tablenotetext{b}{Time center of the MAXI scan referred to the trigger time $T_{\rm trig}$ (= 2011-11-11 05:05:59 UT).}
    \tablenotetext{c}{Unabsorbed flux is in units of erg s$^{-1}$ cm$^{-2}$ in an energy range of $0.7-7.0$ keV.}
    \tablenotetext{d}{The last scan before $T_{\rm trig}$.}
    \tablenotetext{e}{A blackbody model with a temperature of 0.37 keV is assumed (Section~\ref{subsec: upper limit GSC}).}
    \tablenotetext{f}{Upper limit is 90\% confidence.}
    \tablenotetext{g}{The best-fit blackbody model shown in Table~\ref{table: maxi spectrum}.}
    \tablenotetext{h}{The Blackbody + Mekal model with parameters shown in 
      Section~\ref{sec: MAXI/SSC spec M+2 ph} is assumed.}
    \tablenotetext{}{$^{\ast}$We set this time $t_{-1}$. $^{\dagger}$$t_{1}$. $^{\ddagger}$$t_{2}$. $^{\S}$$t_{3}$.}
  \end{center}
\end{table*}

{\it Swift} XRT performed follow-up observations from 0.44~days after $T_{\rm trig}$
\citep[See Table 1 of][]{Li+2012}. {\it Swift} UVOT also observed the optical counterpart
in image mode.
{\it Swift} UVOT grism observations were performed
on 2011 November 19 (+8.23 days after $T_{\rm trig}$)
and 2012 September 30 (324 days).

A ground-based optical spectrum, with relatively high resolution,
was obtained on 2012 May 19 (190 days after the $T_{\rm trig}$)
with the RC spectrograph\footnote{http://www.ctio.noao.edu/spectrographs/60spec/60spec.html}
on the SMARTS
\footnote{The Small and Moderate Aperture Research Telescope System
is a partnership that has overseen operations of 4 small telescopes at
Cerro Tololo Interamerican Observatory since 2003.}/CTIO 1.5m telescope;
this is a long slit spectrograph oriented east-west
\citep{Walter+2004,Walter+2012}.
We used a 1~arcsec slit width and a Loral 1K CCD for the detector.

\section{Analysis and results}

\subsection{Data analysis of MAXI GSC}\label{sec: MAXI/GSC}

On 2011 November 11, the position of MAXI J0158$-$744 was visible by 
three cameras of GSC-H 
\citep[GSC\_2, GSC\_7 and GSC\_8;][]{Mihara+2011,Sugizaki+2011}.
One of these cameras (GSC\_2) was operated at the nominal high voltage ($= 1650$V),
while the other two (GSC\_7 and GSC\_8) were operated at the
reduced voltage ($= 1550$V).
We analyzed the GSC event data version 1.0 or later,
which included the data taken by cameras operated at the nominal and reduced voltages.
In these versions, the position and energy responses of the anodes \#1 and \#2 were significantly
improved from the previous versions (0.x).
We therefore used events taken from all anodes.

To make light curves within the interval of the scan-ID M+0 (Table~\ref{table: maxi obs}),
we followed the method shown in \citet{Morii+2011b}.
Here we selected events within 5~mm of the position 
coincident with this source along the anode wires, which corresponds to
about 2$^\circ$ on the sky.
The obtained light-curve data in energy bands of
$2 - 4$, $4 - 10$, $10 - 20$, and $2 - 20$ keV
were fitted with 
a model consisting of a triangular transit response curve
for a point source with a constant flux and a constant background.
The light curves are consistent with the model at the 90\% confidence level,
meaning that there was no significant variation of the source flux during that scan.

\begin{figure*}
  \begin{center}
    \includegraphics[width=10cm, angle=-90]{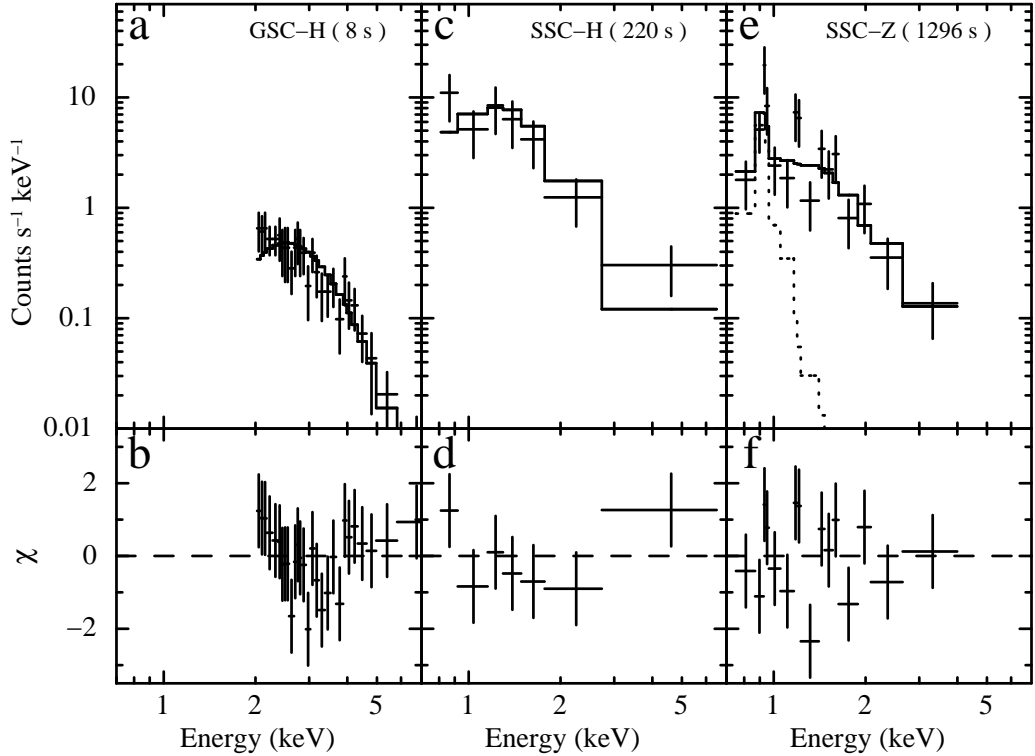}
  \end{center}
  \caption{Energy spectra of MAXI J0158$-$744 observed by MAXI.
    ({\bf Top left})
    Crosses are the GSC-H spectrum at the scan of $+8$~s.
    The histogram is the best-fit blackbody model
    (Table~\ref{table: maxi spectrum}).
    ({\bf Top middle}) Crosses are the SSC-H spectrum at the scan of $+220$~s.
    The histogram is the best-fit blackbody model
    (Table~\ref{table: maxi spectrum}).
    ({\bf Top right}) Crosses are the SSC-Z spectrum at the scan of $+1296$~s.
    The histogram is the best-fit blackbody + Mekal model with parameters shown
    in Section~\ref{sec: MAXI/SSC spec M+2 ph}.
    The Mekal component is shown by a dotted histogram.
    All the spectra are plotted binned with a minimum of 5 counts
    per energy bin.
    Backgrounds are subtracted.
    ({\bf Bottom}) The residuals of the data from the models.
    Error bars, $1\sigma$.}\label{fig: maxi spec}
\end{figure*}

For the spectral analysis,
we removed the GSC\_8 data due to its poor response in the soft energy band.
We selected a concentric circle and
annulus centered at the target as the source and background regions, respectively.
The radius of the source circle was set to 1.8 deg.
The inner and outer radii of the background annulus 
were set to 2.2 and 3.5 deg, respectively.
In both these regions, we excluded a circular region with a radius of 1.5$^\circ$
centered at a near-by bright X-ray source,
SMC X-1 (Fig.~\ref{fig: maxi_finding_chart}b).
The spectrum and response files were
made by the method described in \cite{Nakahira+2011}.
The energy spectra obtained by the GSC Scan-ID M+0
are shown in Fig.~\ref{fig: maxi spec} (left).
We rebinned the data with a minimum of 1 count per energy bin and 
applied Cash statistics \citep{Cash_1979} in the fit.
We used {\tt XSPEC} v12.7.1 for the spectral analysis.

Since the location of this source is near the SMC,
the interstellar absorption by
the total Galactic H I column density towards this source, $N_H$,
and optical extinction $E(B - V)$ are expected to be small.
Thus, we decided to fix them for the following X-ray
and optical spectral analysis.
Two plausible different $N_H$ values are
obtained from the HEASARC web site
\footnote{http://heasarc.gsfc.nasa.gov/cgi-bin/Tools/w3nh/w3nh.pl}:
$1.36 \times 10^{21}$ cm$^{-2}$ by using LAB map \citep{Kalberla+2005} and 
$4.03 \times 10^{20}$ cm$^{-2}$ by using DL map \citep{Dickey_Lockman_1990}.
The corresponding optical extinctions $E(B - V)$
are derived to be 0.28 and 0.084 mag, respectively,
by using the relation with the H I column density
\citep{Bohlin_Savage_Drake_1978}.
On the other hand, the map of dust infrared emission
\citep{Schlegel_Finkbeiner_Davis_1998} suggests
$E(B - V) = 0.050$, which is closer to that from the DL map.
Therefore, we decided to use the latter $N_H$ value,
$4.03 \times 10^{20}$ cm$^{-2}$,
for the interstellar absorption.
In the following, unabsorbed flux is corrected only for
the interstellar absorption.

We fit the GSC X-ray spectrum with absorbed blackbody, power-law,
thermal bremsstrahlung
and Mekal \citep{Mewe_Gronenschild_van-den-Oord_1985} models
from $2.0 - 10.0$ keV
with $N_H$ fixed to $4.03 \times 10^{20}$ cm$^{-2}$;
the results are shown in Table~\ref{table: maxi spectrum}.
The spectrum is statistically consistent with all the models.
Adopting the $N_H$ value of LAB map 
increases the unabsorbed flux by 2\% from that using DL map.
However, the difference in the spectral parameters and unabsorbed flux
are negligibly small, when they are compared with the statistical uncertainty.

\begin{deluxetable}{lcccccccc}
  \tabletypesize{\scriptsize}
  \rotate
  \tablecaption{Spectral fitting parameters with 1$\sigma$ errors
    for the MAXI scans.\label{table: maxi spectrum}}
  \tablewidth{0pt}
  \tablehead{
    \colhead{Model $^a$} &  \colhead{$\Gamma$$^b$} & \colhead{$kT$$^c$} &
    \colhead{$R_{\rm BB}$$^d$} & \colhead{${\it EM}$$^e$} & \colhead{\texttt{abund}$^f$} &
    \colhead{Flux $^g$} & \colhead{Luminosity $^h$} &  \colhead{C-stat$^i$} \\
    \colhead{} &  \colhead{} & \colhead{(keV)} &
    \colhead{($\times 10^{3}$ km)} & \colhead{($\times 10^{63}$ cm$^{-3}$)} & \colhead{} &
    \colhead{(erg s$^{-1}$ cm$^{-2}$)} & \colhead{(erg s$^{-1}$)} &  \colhead{(DOF$^j$)}
  }
  \startdata
  \multicolumn{9}{c}{MAXI GSC-H (Scan-ID M+0, $+8$ s)} \\
    \hline
    PL      &  $4.89^{+0.29}_{-0.28}$   &  \dots &   \dots    &  \dots  & \dots  &
    $6.92^{+0.55}_{-0.52} \times 10^{-9}$ & $2.98^{+0.24}_{-0.22} \times 10^{39}$  &  43.4 (60) \\
    BB      &   \dots   &  $0.45^{+0.03}_{-0.03}$  & $1.26^{+0.29}_{-0.23}$ &  \dots & \dots  &
    $6.74^{+0.27}_{-1.01} \times 10^{-9}$ & $2.90^{+0.12}_{-0.43} \times 10^{39}$  &  51.8 (60)  \\
    TB      &   \dots   &  $0.93^{+0.10}_{-0.09}$ &  \dots & $5.9^{+2.1}_{-1.5}$ & \dots  &
    $6.84^{+0.31}_{-0.98} \times 10^{-9}$ & $2.95^{+0.13}_{-0.42} \times 10^{39}$ &  45.3 (60) \\
    Mekal   &   \dots   & $0.94^{+0.08}_{-0.11}$ &  \dots &  $4.3^{+1.6}_{-0.9}$  & 0.1 (fix) &
    $6.80^{+0.38}_{-0.67} \times 10^{-9}$ & $2.93^{+0.17}_{-0.29} \times 10^{39}$ &  44.6 (60) \\
    Mekal   &   \dots   & $0.97^{+0.11}_{-0.10}$ &  \dots &  $2.4^{+3.0}_{-2.1}$ & $0.50^{+5.74}_{-0.50}$ &
    $6.74^{+0.10}_{-5.14} \times 10^{-9}$ & $2.90^{+0.04}_{-2.21} \times 10^{39}$ & 44.2 (59) \\
    \hline
    \multicolumn{9}{c}{MAXI SSC-H (Scan-ID M+1, $+220$ s)} \\
    \hline
    PL      &   $2.85^{+0.34}_{-0.32}$ &  \dots  &   \dots   &  \dots & \dots  &
    $4.96^{+0.84}_{-0.84} \times 10^{-8}$ & $2.13^{+0.36}_{-0.36} \times 10^{40}$  & 20.5 (36) \\
    BB      &   \dots   & $0.37^{+0.05}_{-0.05}$ &  $4.8^{+1.6}_{-1.2}$  &   \dots  & \dots  &
    $4.16^{+0.47}_{-1.20} \times 10^{-8}$ & $1.79^{+0.20}_{-0.52} \times 10^{40}$ & 30.2 (36) \\
    TB      &   \dots   & $1.46^{+0.50}_{-0.35}$  & \dots &  $4.5^{+1.7}_{-1.2}$  & \dots  &
    $4.71^{+0.56}_{-1.13} \times 10^{-8}$ & $2.03^{+0.24}_{-0.49} \times 10^{40}$ & 22.5 (36) \\
    Mekal   &   \dots   & $1.61^{+0.48}_{-0.36}$ &  \dots &  $3.4^{+0.7}_{-0.6}$    & 0.1 (fix)  &
    $4.71^{+0.64}_{-0.87} \times 10^{-8}$ & $2.03^{+0.28}_{-0.37} \times 10^{40}$ & 24.0 (36) \\
    Mekal   &   \dots   & $1.51^{+0.44}_{-0.41}$ &  \dots &  $4.1^{+1.5}_{-1.0}$ & $ < 0.25$(90\%) &
    $4.77^{+1.36}_{-0.83} \times 10^{-8}$ & $2.05^{+0.59}_{-0.36} \times 10^{40}$ & 22.6 (35) \\
    \hline
    \multicolumn{9}{c}{MAXI SSC-Z (Scan-ID M+2, $+1296$ s)} \\
    \hline
    PL      &  $3.19^{+0.26}_{-0.24}$ &  \dots   &  \dots    &  \dots  & \dots  &
    $1.71^{+0.20}_{-0.18} \times 10^{-8}$  & $7.37^{+0.87}_{-0.78} \times 10^{39}$  & 83.4 (69) \\
    BB      &   \dots   & $0.33^{+0.03}_{-0.03}$ &  $2.29^{+0.48}_{-0.39}$ &  \dots & \dots  &
    $1.50^{+0.10}_{-0.28} \times 10^{-8}$  & $6.48^{+0.45}_{-1.21} \times 10^{39}$ &  84.2 (69) \\
    TB      &   \dots   & $0.94^{+0.17}_{-0.14}$ &  \dots &  $2.61^{+0.71}_{-0.54}$ & \dots  &
    $1.61^{+0.12}_{-0.30} \times 10^{-8}$  & $6.91^{+0.50}_{-1.30} \times 10^{39}$  &  80.8 (69) \\
    Mekal   &   \dots   & $1.03^{+0.10}_{-0.09}$ &  \dots &  $1.42^{+0.18}_{-0.16}$  & 0.1 (fix) &
    $1.57^{+0.17}_{-0.18} \times 10^{-8}$  & $6.75^{+0.73}_{-0.77} \times 10^{39}$  &  80.8 (69) \\
    Mekal   &   \dots   & $0.96^{+0.13}_{-0.10}$ &  \dots &  $1.86^{+0.57}_{-0.46}$ & 
    $0.041^{+0.058}_{-0.031}$ &
    $1.58^{+0.11}_{-0.45} \times 10^{-8}$  & $6.80^{+0.49}_{-1.93} \times 10^{39}$  &  79.4 (68) \\
    \tableline
    \enddata
    %
    \tablenotetext{a}{Absorbed power-law(PL), blackbody(BB), thermal bremsstrahlung(TB) and
      Mekal \citep{Mewe_Gronenschild_van-den-Oord_1985} models are applied
      with $N_H$ fixed to $4.03 \times 10^{20}$~cm$^{-2}$ (Section \ref{sec: MAXI/GSC}).}
    \tablenotetext{b}{Photon index.}
    \tablenotetext{c}{Temperature.}
    \tablenotetext{d}{Blackbody radius at the distance of 60~kpc.}
    \tablenotetext{e}{Emission measure (${\it EM} = \int n_e n_p dV$) at the distance of 60~kpc,
      where $n_e$ and $n_p$ are electron
      and proton number densities, respectively.}
    \tablenotetext{f}{The abundance parameter (\texttt{abund})
      of Mekal model in XSPEC equals to $10^{[X/H]}$,
      where $[X/H]$ is a metallicity of a metal $X$
      \citep{Mewe_Gronenschild_van-den-Oord_1985,Anders_Grevesse_1989}.}
    \tablenotetext{g}{Unabsorbed flux at $2.0 - 10.0$~keV for GSC and $0.7 - 7.0$~keV for SSC.}
    \tablenotetext{h}{{L}uminosity at $2.0 - 10.0$~keV for GSC and $0.7 - 7.0$~keV for SSC
      at the distance of 60 kpc.}
    \tablenotetext{i}{Cash statistics \citep{Cash_1979} calculated
      by binning the data with a minimum of 1 count per energy bin.}
    \tablenotetext{j}{Degrees of freedom.}
\end{deluxetable}

\subsection{Data analysis of MAXI SSC} \label{sec: MAXI/SSC}

After the first detection of MAXI J0158$-$744 with the MAXI GSC scan at $+8$~s,
MAXI SSC detected this source twice as shown in Table~\ref{table: maxi obs}.
Before $+8$~s, MAXI J0158$-$744 was below the SSC detection limit of 
$\sim$200~mCrab
in each night scan.
For the spectral analysis, we selected the source regions 
as shown in Fig.~\ref{fig: maxi_finding_chart}c and
reduced the SSC events in the same way as \cite{Kimura+2012}.
While the second SSC detection at $+1296$~s was done at night in the orbit,
the first SSC at $+220$~s was done at the day-time.
Since the SSC day-time data were contaminated by the visible/infrared light
from the Sun \citep{Tsunemi+2010},
we have to be careful of the analysis of the $+220$~s scan data.
We estimated the area suffering from the contamination 
based on the event distribution,
which led to the decision that
63\% of the source area was not suitable for the spectral analysis.
We thus used data from the remaining 37\% area in the analysis.

The energy spectra obtained by the SSC scans at $+220$~s and $+1296$~s are
shown in Fig.~\ref{fig: maxi spec} (middle, right).
In the latter spectrum, since emission lines seemed to be present,
we analyzed the SSC spectra with special care by following 
the method shown in ``low count spectra'' of the XSPEC wiki site
\footnote{https://astrophysics.gsfc.nasa.gov/XSPECwiki/low\_count\_spectra}.
To avoid losing information on emission lines due to the spectral binning,
we did not bin the data and applied Cash statistics \citep{Cash_1979} 
to the fits.
We fit the source spectra with a model consisting of
a source and a background component.
Here, the background model was analytically-described to approximate
the actual background spectrum in $0.7 - 7.0$~keV.
It was constructed by averaging the two year SSC data weighted
with geomagnetic cut-off-rigidity,
because the background events are caused by charged particles and cosmic diffuse X-rays.
The details of the background spectrum are shown in \citet{Kimura+2012}.
We fit the spectra by absorbed blackbody, power-law, thermal bremsstrahlung and
Mekal \citep{Mewe_Gronenschild_van-den-Oord_1985} models in $0.7 - 7.0$~keV
with $N_H$ fixed to $4.03 \times 10^{20}$~cm$^{-2}$ (Section \ref{sec: MAXI/GSC})
and the results are shown in Table~\ref{table: maxi spectrum}.
For the Mekal model we let the abundance parameter 
(hereafter, we abbreviate it to \texttt{abund}.)
\footnote{The abundance parameter (\texttt{abund}) of Mekal model in XSPEC
equals to $10^{[X/H]}$.
Here, $[X/H] = \log_{10}\left( n_{\rm X} / n_{\rm H} \right)_{\rm source}
- \log_{10}\left( n_{\rm X} / n_{\rm H} \right)_{\rm sun}$
is a metallicity of a metal $X$, where $\left( n_{\rm X} \right)_{\rm Y}$
represents the number density of an element ($X$) in a source ($Y$).}
be free or fixed to 0.1
\citep[a typical abundance of the SMC;][]{Carrera+2008}.
For the first SSC spectrum ($+220$~s), the data are statistically consistent with all the models,
while for the second, the free abundance Mekal model is preferred.
Adopting the $N_H$ value of LAB map increases the unabsorbed fluxes
by up to 20\% from those using DL map. However,
differences in all the spectral parameters and unabsorbed fluxes are
not significant (less than 2.6 sigma level of the statistical uncertainty).

As shown in Table~\ref{table: maxi spectrum},
the initial X-ray outburst of MAXI J0158$-$744 detected by MAXI GSC and SSC
was peaked at $+220$~s. The peak luminosity was extraordinarily luminous,
$2 \times 10^{40}$ erg s$^{-1}$ in $0.7-7.0$ keV,
which is two orders of magnitude larger than
the Eddington luminosity of a solar mass object.
In the following subsections,
we investigate the MAXI SSC spectrum at $+1296$~s,
where emission lines are apparent,
with two scenarios: shock-induced emission
and photospheric emission at the fireball phase
(See Section~\ref{sec: discussion}).

\subsubsection{Detailed spectral analysis of MAXI SSC at $+1296$~s
with a shock-induced emission model}\label{sec: MAXI/SSC spec M+2}

\begin{figure*}
  \begin{tabular}{cc}
    \includegraphics[width=5.5cm, angle=-90]{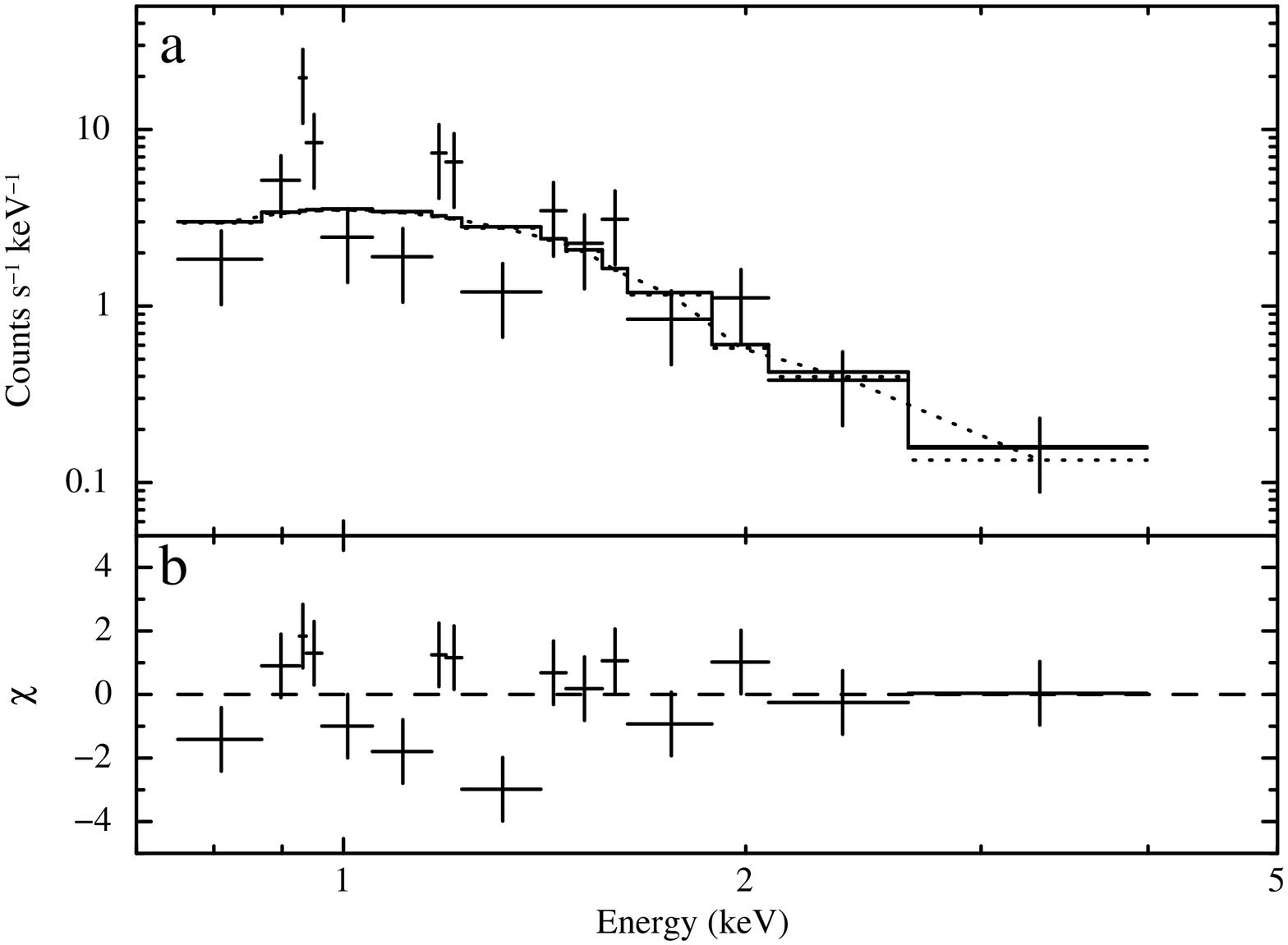}
    &
    \includegraphics[width=5.5cm, angle=-90]{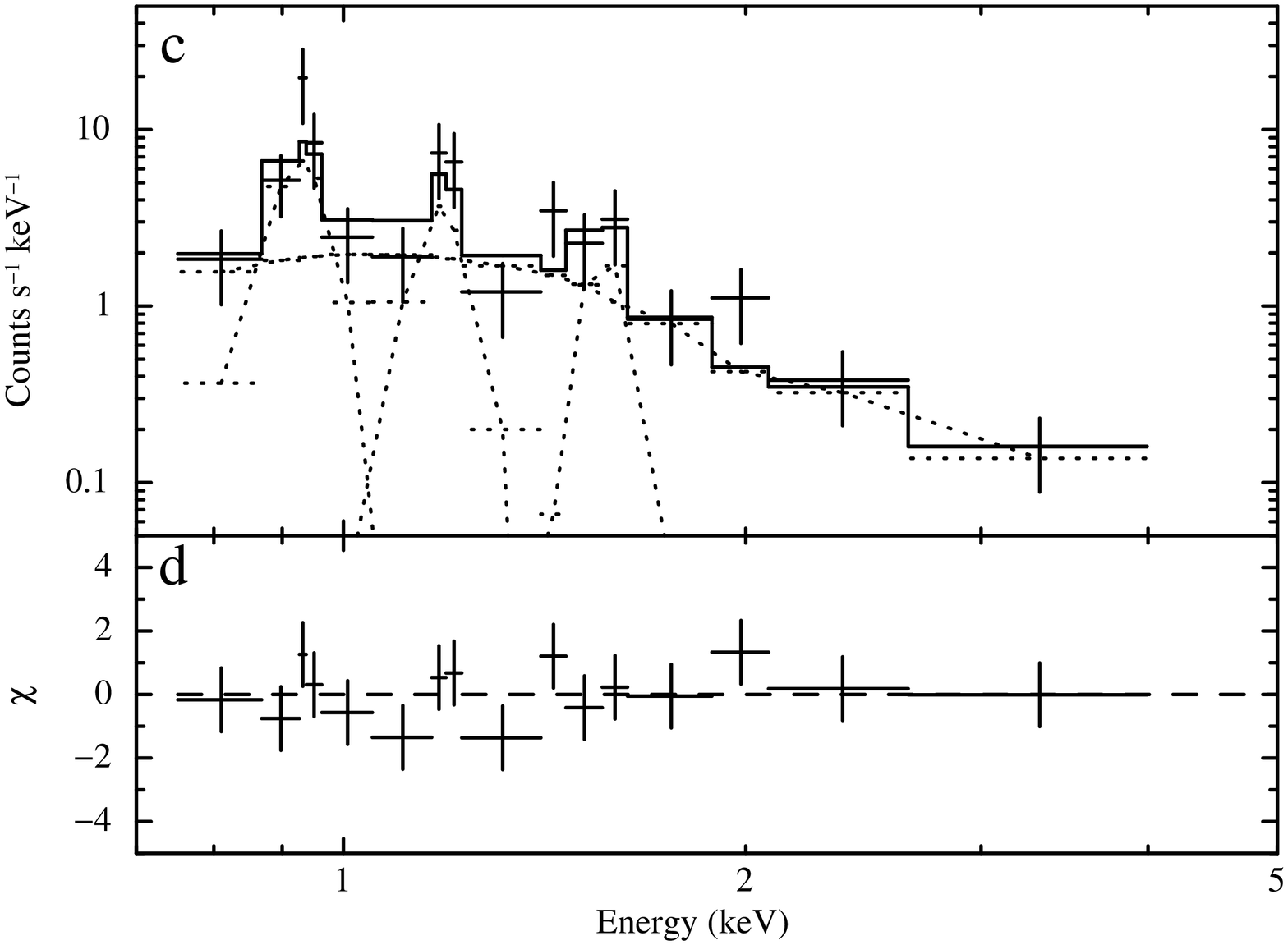}
    \\

    \includegraphics[width=5.5cm, angle=-90]{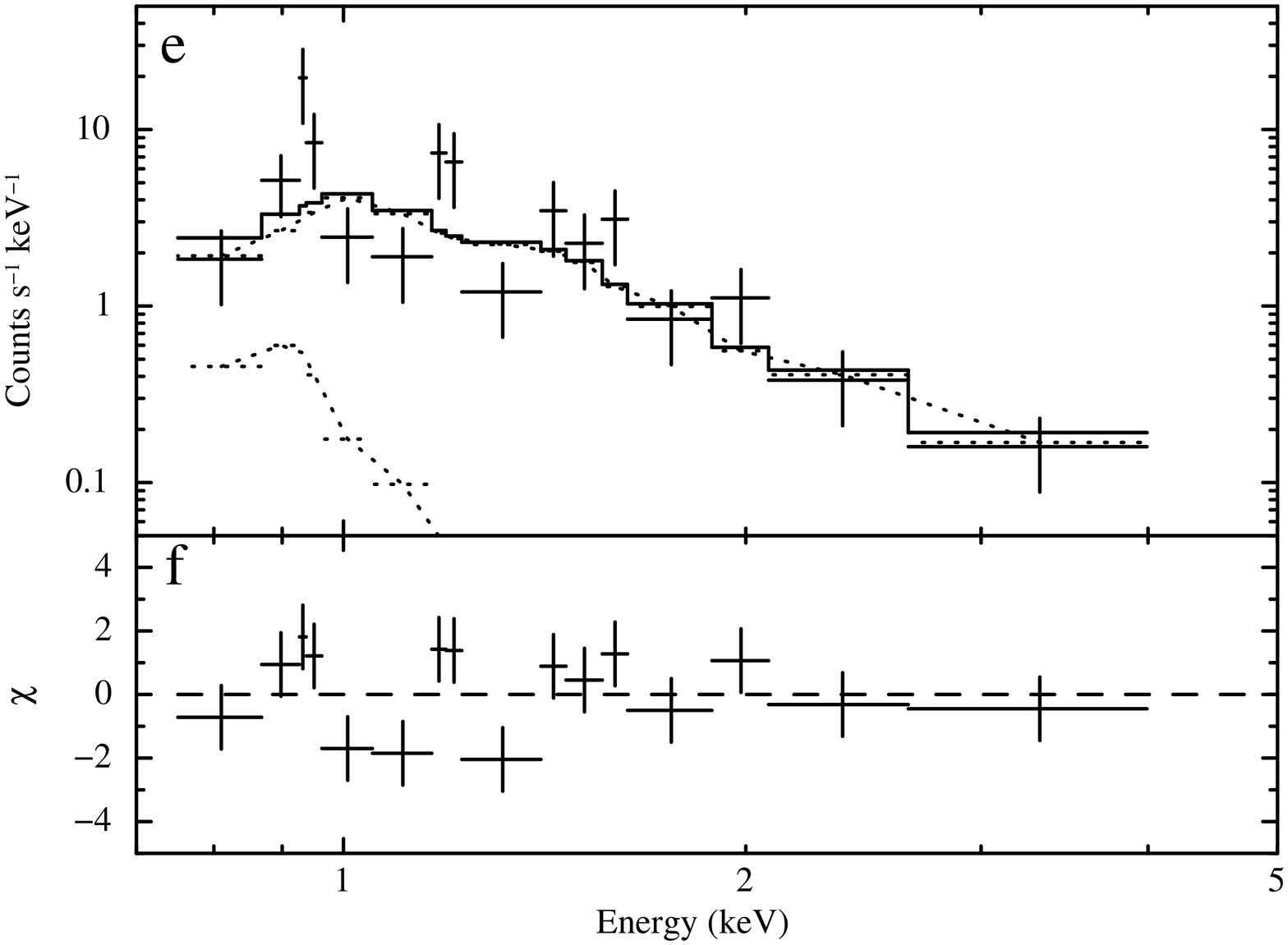}
    &
    \includegraphics[width=5.5cm, angle=-90]{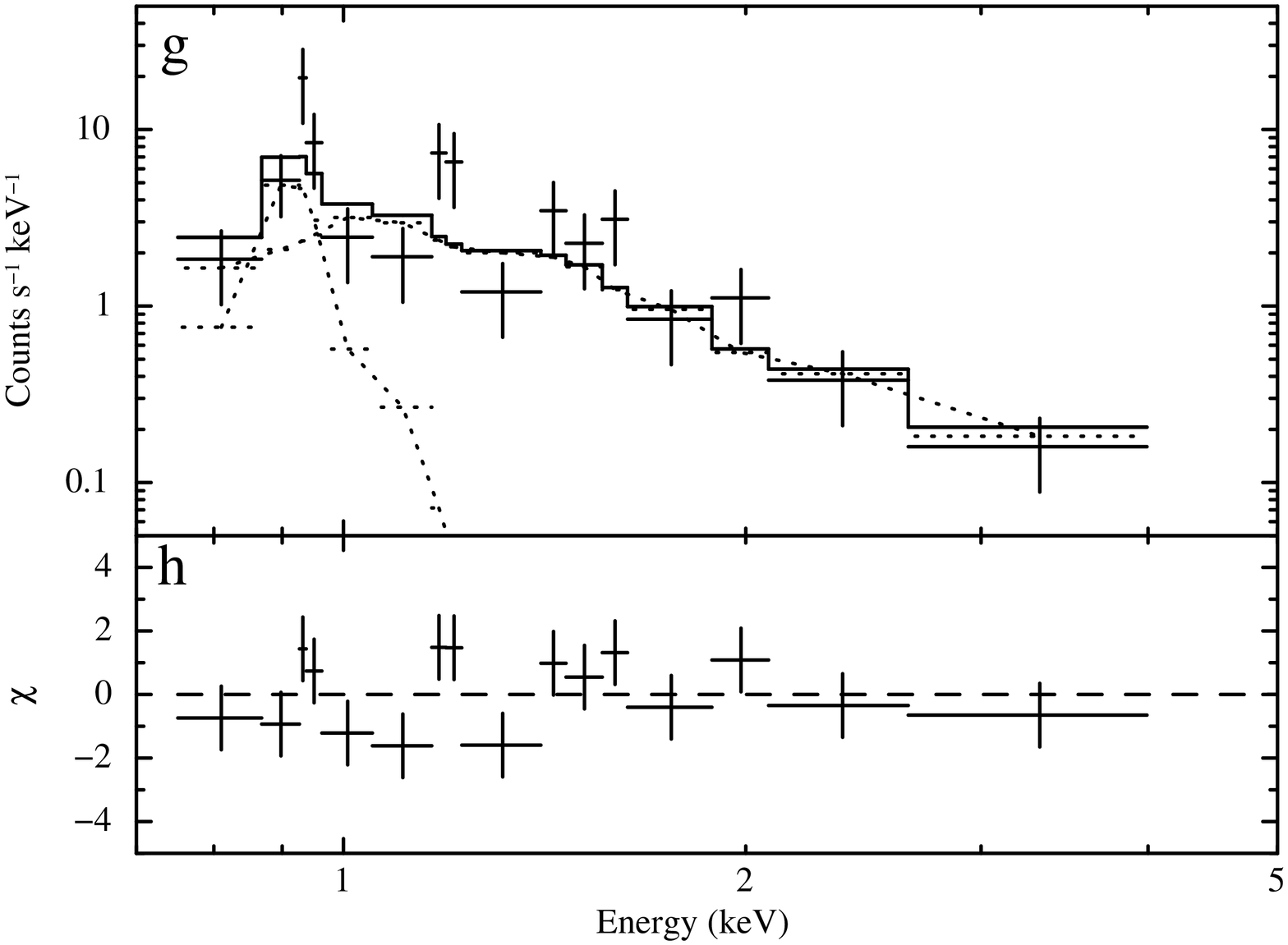}
    \\

  \end{tabular}
  \caption{MAXI SSC-Z spectrum of MAXI J0158$-$744 at $+1296$~s.
    ({\bf a, c}) The histograms are the best-fit thermal bremsstrahlung
    ({\bf a}) and thermal bremsstrahlung + 3 Gaussian lines models ({\bf c}).
    ({\bf e}) The histogram is the best-fit
    Mekal$_{\rm LT}$ + Mekal$_{\rm HT}$ model,
    where the \texttt{abund}s of Ne and the other elements
    in the Mekal$_{\rm LT}$ component
    are set to 0.1 and 0.1, respectively
    (Table~\ref{table: vmekal}, the first row).
    ({\bf g}) The histogram is the same model as {\bf e},
    where these \texttt{abund}s are set to 10.0 and 0.1
    (Table~\ref{table: vmekal}, the fifth row).
    All the spectra are plotted binned with a minimum of 5 counts
    per energy bin.
    ({\bf b, d, f, h}) The residuals of the data from the models.
    Error bars, 1~$\sigma$.
    \label{fig: m002_various_fits}
  }
\end{figure*}

To investigate the emission lines in the spectrum at $+1296$~s,
we first fit the spectrum
with models consisting of thermal bremsstrahlung continuum and Gaussian lines,
whose widths were fixed to be small against
the detector energy resolution, 85~eV (FWHM) at 1.0~keV \citep{Kimura+2012}.
The best-fit parameters are summarized
in Table~\ref{table: M+2 spectrum} (upper) and
the models are shown in Fig.~\ref{fig: m002_various_fits} (abcd).
The results of the likelihood ratio tests \citep{Cash_1979}
in the last two rows of Table~\ref{table: M+2 spectrum} (upper)
indicate that 
the addition of the Gaussian lines at the energies $E_1$, $E_2$ and $E_3$ one by one
improves the fits
with a chance probability of 0.0044, 0.084 and 0.085, respectively.
The line at the energy $E_1$ is the most significant, 
and is inferred to be a resonance line of He-like neon (0.922~keV).
The other two lines are less significant than $2\sigma$
and no corresponding major lines exist at these energies.
However, the line center energy $E_2$ may suggest 
a radiative recombination continuum of He-like neon (1.20~keV) or
Lyman beta line of H-like neon (1.24~keV),
and the $E_3$ may suggest
a resonance line of He-like aluminum (1.60~keV) or
$1s3p$ $^{1}P_{1}$ $\rightarrow$ $1s^{2}$ $^{1}S_{0}$ line of He-like magnesium (1.58~keV).
The neon emission line suggests that
the initial bright outburst would have been produced
by an optically-thin thermal emission mechanism,
whose site was a region heated by the shock wave of a nova explosion
as seen in some novae 
\citep[RS Ophiuchi and V407 Cyg;][]{Sokoloski+2006, Nelson+2012}.

\begin{table*}
  \begin{center}
  \caption{Spectral fitting parameters of a continuum + emission lines
    for the SSC-Z scan at $+1296$~s and the 
    likelihood ratio test.\label{table: M+2 spectrum}}
  \begin{tabular}{lcccc}
    \tableline \tableline
    Model      &  
    TB$^a$          &
    TB + Line$^b$       &  TB + 2 Lines       &
    TB + 3 Lines             \\
    \tableline
    $kT$$^c$ (keV) & 
                   $0.94^{+0.17}_{-0.14}$ &
                   $1.12^{+0.25}_{-0.19}$ & $1.21^{+0.34}_{-0.20}$ &
                   $1.28^{+0.36}_{-0.25}$ \\
      ${\it EM}$$^d$ ($\times 10^{63}$ cm$^{-3}$)  &
                   $2.61^{+0.71}_{-0.54}$  &
                   $1.78^{+0.53}_{-0.40}$  & $1.42^{+0.44}_{-0.37}$ &
                   $1.17^{+0.43}_{-0.32}$  \\
      $E_1$$^e$ (keV) &
                   \dots                  &
                   $0.93^{+0.01}_{-0.01}$  & $0.93^{+0.01}_{-0.01}$ &
                   $0.93^{+0.01}_{-0.01}$ \\
      $EW_1$$^f$ (keV)   &
                   \dots                  &
                   $0.19^{+0.13}_{-0.07}$  & $0.26^{+0.22}_{-0.10}$ &
                   $0.33^{+0.26}_{-0.11}$   \\
      $E_2$$^g$ (keV) &
                     \dots                 &
                     \dots                 & $1.19^{+0.02}_{-0.02}$ &
                   $1.19^{+0.02}_{-0.02}$ \\
      $EW_2$ (keV)   &
                     \dots                 &
                     \dots                 & $0.14^{+0.14}_{-0.08}$ &
                   $0.18^{+0.20}_{-0.08}$  \\
      $E_3$$^h$ (keV) &
                     \dots                 &
                     \dots                 &   \dots                &
                   $1.57^{+0.03}_{-0.04}$ \\
      $EW_3$ (keV)   &
                     \dots                 &
                     \dots                 & \dots  &
                   $0.24^{+0.26}_{-0.14}$  \\
      $C$-stat$^i$(DOF$^j$)      &
                    342.9(1723)      &
                    332.1(1721)      &  327.1(1719)    &
                    322.2(1717)           \\
    \tableline
      $\Delta C$($\Delta$DOF)$^k$ &
                     \dots                 &
                     10.8(2)              &  5.0(2)     &
                      4.9(2)              \\
      $P$-value$^l$           &
                     \dots                    &
                      0.0044  &   0.084     &
                      0.085   \\
      \tableline \tableline
      Model      &  
                     BB$^m$          &
                     BB + Line       &  BB + 2 Lines       &
                     BB + 3 Lines             \\
      \tableline
      $kT$$^c$ (keV) & 
                   $0.33^{+0.03}_{-0.03}$ &
                   $0.38^{+0.04}_{-0.03}$ & $0.41^{+0.04}_{-0.04}$ &
                   $0.42^{+0.05}_{-0.04}$  \\
      $R_{\rm BB}$$^n$ ($\times 10^{3}$ km)  &
                   $2.29^{+0.48}_{-0.39}$ &
                   $1.54^{+0.35}_{-0.28}$ & $1.22^{+0.35}_{-0.23}$ &
                   $1.10^{+0.35}_{-0.22}$  \\
      $E_1$$^e$ (keV) &
                   \dots                   &
                   $0.93^{+0.01}_{-0.01}$  & $0.93^{+0.01}_{-0.01}$ &
                   $0.93^{+0.01}_{-0.01}$ \\
      $EW_1$$^f$ (keV)   &
                   \dots                  &
                   $0.32^{+0.21}_{-0.11}$ & $0.45^{+0.36}_{-0.14}$ &
                   $0.55^{+0.45}_{-0.18}$ \\
      $E_2$$^g$ (keV) &
                     \dots                 &
                     \dots                 & $1.19^{+0.02}_{-0.02}$ &
                   $1.19^{+0.02}_{-0.02}$ \\
      $EW_2$ (keV)   &
                     \dots                 &
                     \dots                 & $0.16^{+0.16}_{-0.08}$ &
                   $0.20^{+0.21}_{-0.09}$  \\
      $E_3$$^h$ (keV) &
                     \dots                 &
                     \dots                 &   \dots                &
                   $1.57^{+0.04}_{-0.05}$ \\
      $EW_3$ (keV)   &
                     \dots                 &
                     \dots                 & \dots  &
                   $0.16^{+0.22}_{-0.11}$  \\
      $C$-stat$^i$(DOF$^j$)      &
                    343.8(1723)      &
                    326.7(1721)      &  321.1(1719)    &
                    318.2(1717)           \\
    \tableline
      $\Delta C$($\Delta$DOF)$^k$ &
                     \dots                 &
                     17.1(2)              &  5.6(2)     &
                      2.9(2)              \\
      $P$-value$^l$           &
                     \dots                    &
                      $1.9 \times 10^{-4}$  &  0.061 &
                      0.23  \\
    \tableline
  \end{tabular}
  \tablenotetext{a}{Thermal bremsstrahlung.}
  \tablenotetext{b}{Gaussian line.}
  \tablenotetext{c}{Temperature.}
  \tablenotetext{d}{Emission measure.}
  \tablenotetext{e}{{L}ine energy ($E_1$) is constrained
    between 0.7 and 1.1~keV.}
  \tablenotetext{f}{Equivalent width.}
  \tablenotetext{g}{$E_2$: between 1.1 and 1.3 keV.}
  \tablenotetext{h}{$E_3$: between 1.3 and 1.7~keV.}
  \tablenotetext{i}{Cash statistics \citep{Cash_1979} without binning.}
  \tablenotetext{j}{Degrees of freedom.}
  \tablenotetext{k}{The difference of the $C$-stat (DOF)
    between this column and
    the next column to the left.}
  \tablenotetext{l}{$\Delta C$ is distributed as $\chi^2$
    with degrees of freedom
    of $\Delta$DOF \citep{Cash_1979}.}
  \tablenotetext{m}{Blackbody.}
  \tablenotetext{n}{Blackbody radius at the distance of 60 kpc.}
  \end{center}
\end{table*}

We next tried to fit the spectrum with more physically motivated models.
Although it can be fit with an optically-thin thermal emission model
(Mekal in {\tt XSPEC} terminology)
with a temperature of $\sim 1.0$~keV (Table~\ref{table: maxi spectrum}),
the model cannot
produce the observed strong He-like neon line.
To reproduce the He-like neon, another optically-thin thermal component
with a lower temperature (about 0.1~keV) is necessary.
Thus, we examined a model consisting of two Mekal models
(Mekal$_{\rm LT}$ + Mekal$_{\rm HT}$), 
whose temperatures are $\sim 0.1$~keV in the lower component (LT) and
$\sim 1.0$~keV in the higher component (HT), respectively.
The best-fit result in Table~\ref{table: maxi spectrum} shows
that the \texttt{abund} of the Mekal$_{\rm HT}$
is consistent with that of the SMC.
This conclusion, however, is not completely correct, because the fit included an energy range 
affected by the He-like neon line produced by the Mekal$_{\rm LT}$ component.
To determine the \texttt{abund} of Mekal$_{\rm HT}$,
we fit the spectrum excluding the energy range $0.8-1.0$~keV,
and then obtained an upper limit for the \texttt{abund} of 0.25 (90\% confidence limit),
which is consistent with that of the SMC \citep{Carrera+2008}.
We thus decided to fix the \texttt{abund} of the Mekal$_{\rm HT}$ component to 0.1.
For the Mekal$_{\rm LT}$ component, the observed strong He-like neon line suggests
a large abundance for neon.
We postulate that the Mekal$_{\rm LT}$ component was produced
in a reverse shocked region whose material was ejecta from the nova explosion
(See Section~\ref{sec: discussion}).

To estimate the \texttt{abund}s of the Mekal$_{\rm LT}$ component,
we fit the spectrum with the Mekal$_{\rm LT}$ + Mekal$_{\rm HT}$ model
assuming six combinations of the \texttt{abund}s of neon and the other elements
for Mekal$_{\rm LT}$ as shown in Table~\ref{table: vmekal}.
Here, at the first step in the spectral fit,
we let the temperature and emission measure of the Mekal$_{\rm HT}$
component freely vary.
When the 1$\sigma$ error range of the temperature of the Mekal$_{\rm LT}$ was
not constrained to less than 0.3~keV in the first step
(the first three cases of Table~\ref{table: vmekal}),
we fixed the temperature of Mekal$_{\rm HT}$
and then the emission measure of Mekal$_{\rm HT}$ to the best-fit values.
These best fit values were obtained by fitting 
the same spectrum, excluding the energy range $0.8-1.0$~keV,
with a single Mekal component and the \texttt{abund} fixed to 0.1.
For the first case of Table~\ref{table: vmekal},
the 1$\sigma$ error range of the Mekal$_{\rm LT}$ could not be constrained
to be less than 0.3~keV, even after both the temperature
and emission measure of Mekal$_{\rm HT}$ were fixed.
Figure~\ref{fig: m002_various_fits} (efgh) presents
the difference in these spectral fits
with respect to the neon \texttt{abund}s of the Mekal$_{\rm LT}$ component.
As a result, the neon \texttt{abund} of the Mekal$_{\rm LT}$ was
suggested to be much higher than that of the SMC (Table~\ref{table: vmekal}),
and it indicates that the Mekal$_{\rm LT}$ component originates
in ejecta from the nova.
The unabsorbed flux in $0.7 - 7.0$ keV, assuming the 
Mekal$_{\rm LT}$ + Mekal$_{\rm HT}$ model with parameters shown in the
fifth row of Table \ref{table: vmekal} (the best-fit model),
is $1.63^{+0.19}_{-0.17} \times 10^{-8}$ erg s$^{-1}$ cm$^{-2}$.

\begin{table*}
  \begin{center}
    \caption{Spectral fitting parameters of the Mekal$_{\rm LT}$$^a$ + Mekal$_{\rm HT}$$^b$ model
      for the SSC-Z scan at $+1296$~s with 1$\sigma$ errors.\label{table: vmekal}}
    \begin{tabular}{ll|ccccc}
      \tableline \tableline
      \multicolumn{2}{c|}{\texttt{abund}$^c$}    & $kT$(LT$^d$) &  ${\it EM}$(LT) & 
      $kT$(HT$^e$) &  ${\it EM}$(HT) &  C-stat$^f$ \\ 
      Ne                    & Others                 & (keV)    &  ($\times 10^{63}$ cm$^{-3}$) &
      (keV)    & ($\times 10^{63}$ cm$^{-3}$) & (DOF$^g$) \\
      \tableline
      $0.1$ & $0.1$  &  0.13$^h$(unconstrained)$^i$  &  11.5$^h$(0.0 $-$ 194)$^j$ &
      $1.29$ (fix)$^k$ &  $1.19$ (fix)$^k$ &  81.6 (69) \\

      $1.0$ & $1.0$  &  $0.14^{+0.11}_{-0.05}$       &  $1.83^{+18.87}_{-1.80}$      &
      $1.29$ (fix)$^k$ &  $1.19$ (fix)$^k$ & 80.6 (69) \\ 

      $10.0$ & $10.0$  &   $0.13^{+0.08}_{-0.05}$       &   $0.19^{+1.88}_{-0.18}$  &
      $1.29$ (fix)$^k$ &  $1.19$ (fix)$^k$ & 80.5 (69) \\ 

      $1.0$                 & $0.1$  &    $0.13^{+0.06}_{-0.04}$ &  $10.3^{+184.5}_{-8.3}$  &
      $1.44^{+0.16}_{-0.28}$ &  $1.03^{+0.24}_{-0.12}$ &  74.8 (67) \\ 

      $^*$$10.0$                 & $0.1$  & 0.13$^h$($0.0808^l - 0.24$)$^j$ &  $1.43^{+58.32}_{-1.30}$  &
      $1.44^{+0.17}_{-0.21}$ & $1.02^{+0.20}_{-0.09}$  & 73.8 (67) \\ 

      $10.0$                 & $1.0$  &  $0.13^{+0.06}_{-0.04}$  &   $1.11^{+21.24}_{-0.87}$ &
      $1.42^{+0.17}_{-0.25}$ &  $1.05^{+0.20}_{-0.15}$ &  74.6 (67) \\ 
      \tableline
    \end{tabular}
    \tablenotetext{a}{To set different \texttt{abund}s for neon and the other elements,
      vmekal model in {\tt XSPEC} was used for the Mekal$_{\rm LT}$ component.}
    \tablenotetext{b}{The \texttt{abund} of Mekal$_{\rm HT}$ was fixed to 0.1.}
    \tablenotetext{c}{The \texttt{abund}s of the Mekal$_{\rm LT}$ component.}
    \tablenotetext{d}{{L}T: Lower temperature.}
    \tablenotetext{e}{HT: Higher temperature.}
    \tablenotetext{f}{Cash statistics \citep{Cash_1979} calculated
      by binning the data with a minimum of 1 count per energy bin.}
    \tablenotetext{g}{Degrees of freedom.}
    \tablenotetext{h}{Best-fit value.}
    \tablenotetext{i}{$1\sigma$ error interval is not constrained between 0.0808~keV$^l$ and 0.3~keV.}
    \tablenotetext{j}{$1\sigma$ error interval.}
    \tablenotetext{k}{The best-fit temperature and emission measure obtained by a single
      Mekal fits for the same spectrum excluding the energy range $0.8-1.0$~keV
      with the \texttt{abund} fixed to 0.1.}
    \tablenotetext{l}{Computational lower boundary of the Mekal model.}
    \tablenotetext{}{$^*$The best fit and preferred model.}
  \end{center}
\end{table*}

\subsubsection{Detailed spectral analysis of MAXI SSC at $+1296$~s
with photospheric emission at the fireball phase}
\label{sec: MAXI/SSC spec M+2 ph}

The initial bright outburst detected by MAXI may 
also be explained by photospheric emission
at the ignition phase of a nova explosion, the so-called fireball phase
(See Section~\ref{sec: discussion}).
In this scenario, the main continuum component
in the spectrum of MAXI SSC at $+1296$~s
is blackbody emission, while the emission lines
come from the optically-thin region 
surrounding the photosphere (See Fig.~\ref{fig: nova evolution}
in Section~\ref{sec: discussion}).
Thus, we fit the spectrum with
models consisting of blackbody continuum and Gaussian lines.
The results of the fit are shown
in Table~\ref{table: M+2 spectrum} (lower).
The addition of the Gaussian line at the energy $E_1$
significantly improves the fits
with a chance probability of $1.9 \times 10^{-4}$,
while the other two lines are detected at $< 2\sigma$.
The identifications of these lines are the same as
in Section \ref{sec: MAXI/SSC spec M+2}.
The detection of the neon emission line suggests that the spectrum contains
an optically-thin thermal emission component.
Therefore, this spectrum could be explained by
a composite model of a blackbody and a Mekal with a temperature below 0.3~keV,
and an exceptionally large neon abundance,
similar to the two Mekal models in Section \ref{sec: MAXI/SSC spec M+2}.
When the \texttt{abund}s of neon and the other elements
are set to $10$ and 0.1, respectively,
the resultant best-fit spectral parameters are as follows:
the temperature and emission measure of the Mekal component are
0.14($0.0808 - 0.28$)~keV 
(See footnotes $h$, $j$ and $l$ of Table~\ref{table: vmekal})
and $1.13^{+62.97}_{-1.03} \times 10^{63}$~cm$^{-3}$,
respectively.
The temperature and radius of the blackbody component are
$0.39^{+0.04}_{-0.04}$~keV and $1.47^{+0.37}_{-0.31} \times 10^3$~km,
respectively.

\subsection{Upper limits on other MAXI GSC scans}
\label{subsec: upper limit GSC}

In the scans at $-5530$~s and $+5545$~s (Table~\ref{table: maxi obs}),
MAXI J0158$-$744 was not detected by MAXI GSC.
To calculate the upper limits on these fluxes,
we assumed the best-fit Mekal$_{\rm LT}$ + Mekal$_{\rm HT}$ model
obtained by the MAXI SSC scan at $+1296$~s (Section \ref{sec: MAXI/SSC spec M+2})
and the best-fit blackbody model (the Scan-ID M+1 in Table~\ref{table: maxi spectrum}).
In the former model,
the \texttt{abund} of the Mekal$_{\rm HT}$ and 
the Mekal$_{\rm LT}$ was fixed to 0.1 
except for the neon \texttt{abund} in the Mekal$_{\rm LT}$
fixed to 10 (Table~\ref{table: vmekal}, the fifth row).
The 90\% confidence-level upper limits on
the unabsorbed flux in $0.7 - 7.0$~keV for these scans
are $< 1.94 \times 10^{-9}$ and $< 8.29 \times 10^{-10}$ 
erg~s$^{-1}$~cm$^{-2}$ in the two Mekal model,
and $<2.75 \times 10^{-9}$ and $< 9.60 \times 10^{-10}$
erg~s$^{-1}$~cm$^{-2}$ in the blackbody model.

In addition, in all five scans 
between the scan at $+5545$~s (+0.064~d) and the start of
the {\it Swift} XRT follow-up
(i.e. at +0.128, +0.192, +0.256, +0.320, and +0.385~days),
MAXI J0158$-$744 was not detected by MAXI GSC.
The 90\% confidence-level upper limit
on the unabsorbed flux in the 0.7$-$7.0~keV band
for this period was $3.5 \times 10^{-10}$ erg~s$^{-1}$~cm$^{-2}$,
assuming the former model
and $2.5 \times 10^{-10}$ erg~s$^{-1}$~cm$^{-2}$ for the latter model.

\subsection{Analysis of follow-up observations}\label{sec: follow-up}

\subsubsection{Analysis of {\it Swift} observations}\label{sec: Swift obs}

We analyzed the same {\it Swift} XRT archival data as listed in Table 1 of \cite{Li+2012},
using {\it Swift} software version 3.9, released as part of HEASOFT 6.12.
We extracted the source events from a circle with optimal radii ($47^{\prime\prime}-23^{\prime\prime}$)
and the background from an offset circular region of radius 142$^{\prime\prime}$.
For the data on day 0.54, we excluded events from the inner 5$^{\prime\prime}$ of the PSF
to avoid pile-up.
In the spectral fit, we used the
RMFs (Redistribution Matrix Files)
of swxpc0to12s6\_20010101v013.rmf in PC mode
and swxwt0to2s6\_20010101v014.rmf in WT mode.
The ARFs (Ancillary Response Files) were generated by using the commands {\tt xrtexpomap}
(to create the exposure maps) and {\tt xrtmkarf}.

We fit the {\it Swift} XRT spectra with absorbed blackbody or
Mekal models with the intrinsic column $N_H$ allowed to vary.
These models include two absorption components:
the interstellar absorption $N_H$ fixed
at $4.03 \times 10^{20}$ cm$^{-2}$ (Section \ref{sec: MAXI/GSC})
and intrinsic absorption.
In the Mekal model, the \texttt{abund} was fixed to 0.1,
a typical \texttt{abund} of the SMC
\citep[][See also Section~\ref{sec: MAXI/SSC spec M+2}]{Carrera+2008}.
The unabsorbed flux obtained by the blackbody fits are shown in Figure~\ref{fig: lc}.

We also analyzed the {\it Swift} UVOT data obtained at the same time as {\it Swift} XRT,
using the {\it Swift} software version 3.9, released as part of HEASOFT 6.12.
The image data of each filter, from each observation sequence, i.e., with a given observation ID,
were summed using {\tt uvotimsum}. However, for images taken within 2 days of
the outburst ($b$, $u$ and $w1$ bands) individual exposures were long enough
that summing was not necessary.
Photometry of the source in individual sequences was derived via {\tt uvotmaghist},
using an extraction region of radius 5$^{\prime\prime}$ and a suitable background region. Magnitudes 
are  based on the UVOT photometric system \citep{poole2008:MNRAS383}.
{\tt XSPEC} compatible spectral files for the source were created using 
the same region with {\tt uvot2pha}. 

The $u$ band light curve is shown in Figure~\ref{fig: lc}.
We calculated the absolute magnitude of the enhanced emission after extinction
correction,
where $E(B - V) = 0.050$~mag (Section \ref{sec: MAXI/GSC}),
$A_V/ E(B - V) = 3.1$, $A_u = 1.664 A_V$ \citep{Schlegel_Finkbeiner_Davis_1998},
and the SMC distance of $d = 60$~kpc were assumed.
Here, we subtracted the flux in the plateau phase
($13.58$~mag, average of $u$-band magnitudes from 11.65~days to 27.86~days.).
By fitting it with a linear function,
we obtained the absolute magnitude of $-5.04 \pm 0.07$ mag at 0.44 day
and the speed class indicator parameter defined by
the time to decline 2 mag from maximum \citep{Warner_2008},
$t_{d, 2} = 1.9 \pm 0.2$ ($1\sigma$) days.
This classifies the event 
as a ``very fast nova'' \citep{Warner_2008},
assuming that the optical enhancement was due to
the photospheric emission as in usual novae.

To investigate the optical enhanced emission, we made a difference spectrum 
from the {\it Swift} UVOT photometry over the six filter bandpasses
\footnote{Central wavelengths (FWHM) in Angstroms \citep{poole2008:MNRAS383}:
{\it v}: 5468 (769), {\it b}: 4392 (975), {\it u}: 3465 (785), 
{\it uvw1}: 2600 (693), {\it uvm2}: 2246 (498), {\it uvw2}: 1928 (657).}
between day 1.5 and 149,
and we fit it with a blackbody model
with fixed interstellar extinction (Section \ref{sec: MAXI/GSC}).
The blackbody temperature of  $1.22^{+0.11}_{-0.10} \times 10^{4}$~K
and the radius of $6.58^{+1.03}_{-0.89} \times 10^{11}$~cm
were obtained at the best fit.
We also fit the spectrum from day 149 with the blackbody model,
obtaining a temperature of $2.34^{+0.04}_{-0.04} \times 10^{4}$~K
and a radius of $5.81^{+0.14}_{-0.14} \times 10^{11}$~cm.
The extrapolation of the best-fit blackbody spectrum 
towards the UV region is consistent with the UV flux obtained by
Galaxy Evolution Explorer
\citep[GALEX;][]{Morrissey+2005}
during the pre-outburst phase \citep{Li+2012}.
In addition, \cite{Li+2012} reported that the I-band flux returned to the pre-outburst level
$\sim 60$ days after $T_{\rm trig}$.
So, we can assume that the flux at day 149 contains 
only emission from the binary companion star.
The obtained temperature and radius
are consistent with those of a B-type star,
as shown in \cite{Li+2012}.

{\it Swift} UVOT grism spectra are close to that of
an early B-type star. No clear emission lines can be identified
above the noise, as shown in \cite{Li+2012}.

\subsubsection{Ground-based optical spectroscopy by SMARTS}\label{sec: SMARTS}

We obtained three 200-s spectra of the optical counterpart of MAXI J0158$-$744 
in order to filter for cosmic rays.
We combine the three images, and extract the spectrum by fitting a Gaussian
in the spatial direction at each pixel. Wavelength calibration is accomplished
by fitting a 3$^{rd}$ to 6$^{th}$ order polynomial to the calibration lamp line positions.
The optical spectrum covers nearly the entire optical band ($3300-9500\AA$)
at $17\AA$ resolution.

There are clear emission lines of $H_\alpha$ and $H_\beta$,
with equivalent widths of $16$~$\AA$ and $1$~$\AA$, respectively.
We could not find any other significant emission or absorption lines
above the noise level. The SMARTS spectrum matches the NTT spectrum of \cite{Li+2012},
albeit with worse signal to noise.


\subsection{Historical X-ray fluxes}\label{sec: history}

To investigate the activity of MAXI J0158$-$744 before the discovery,
we searched for previous X-ray observations of the area including
the target position.
This region was observed by the {\it ROSAT} all-sky survey,
{\it XMM-Newton} slew survey and MAXI GSC.
The source was undetected in all these observations.
We calculated the upper limits on the unabsorbed fluxes
in an energy range of $0.7 - 7.0$ keV,
assuming the best-fit Mekal$_{\rm LT}$ + Mekal$_{\rm HT}$ model
(Table~\ref{table: vmekal}, the fifth row;
Outburst Model)
and a typical spectrum in the SSS phase observed by {\it Swift} XRT,
an absorbed blackbody with a temperature of 0.1 keV \citep[SSS Model;][]{Li+2012}.

The {\it ROSAT} All-Sky Survey covered this field, 
with an exposure of 775~s in total
between 1990 September 22 and December 3.
These data provide a PSPC count rate upper limit of $0.14$ counts s$^{-1}$ 
(90\% confidence limit) over $0.1-2.5$ keV;
corresponding to $< 2.1 \times 10^{-13}$ and
$< 6.5 \times 10^{-14}$ ergs cm$^{-2}$ s$^{-1}$
($0.7 - 7.0$ keV),
assuming Outburst Model and SSS Model,
respectively (Table~\ref{table: historical X-ray}).

{\it XMM-Newton} slewed over the source three times on 2006 November,
2007 October, and 2009 November.
We obtained EPIC pn count rate upper limits
of 0.35, 1.5 and 0.50 counts s$^{-1}$ ($2\sigma$ level)
in $0.2-12$ keV, respectively.
The corresponding unabsorbed fluxes ($0.7 - 7.0$ keV)
are shown in Table~\ref{table: historical X-ray}.

We also analyzed the MAXI GSC image in the 4$-$10 keV band integrated for 7
months from 2009 September 1 to 2010 March 31. Applying the same
analysis procedure as used by \cite{Hiroi+2011}, we obtain a 90\%
confidence-level upper limit of 0.10 mCrab.
It corresponds to an unabsorbed flux of 
$< 1.8 \times 10^{-11}$ ergs cm$^{-2}$ s$^{-1}$ ($0.7 - 7.0$ keV),
assuming Outburst Model.

\begin{table*}
  \begin{center}
    \caption{Summary of {\it ROSAT} all-sky survey and 
      {\it XMM-Newton} slew survey observations.\label{table: historical X-ray}}
    \begin{tabular}{ccccc}
      \tableline \tableline
      Date                &  Telescope  & Exp.(s)$^a$ & Flux(Outburst)$^b$         & Flux(SSS)$^b$ \\
      \tableline
      1990-09-22 -- 1990-12-03  & {\it ROSAT} & 775   & $< 2.1 \times 10^{-13}$ & $< 6.5 \times 10^{-14}$ \\
      2006-11-01 10:03:35 &  {\it XMM-Newton} & 8.9   & $< 3.6 \times 10^{-13}$ & $< 7.3 \times 10^{-14}$ \\
      2007-10-28 11:57:17 &  {\it XMM-Newton} & 2.1   & $< 1.6 \times 10^{-12}$ & $< 3.2 \times 10^{-13}$ \\
      2009-11-30 23:41:46 &  {\it XMM-Newton} & 6.1   & $< 5.2 \times 10^{-13}$ & $< 1.1 \times 10^{-13}$ \\
      \tableline
    \end{tabular}
    \tablenotetext{a}{Exposure in units of seconds. For {\it XMM-Newton}, it is corrected for vignetting to
      on-axis equivalent value.}
    \tablenotetext{b}{The upper limit on the unabsorbed flux in a unit of erg s$^{-1}$ cm$^{-2}$
      in an energy range of $0.7-7.0$ keV, where Outburst and SSS Models are assumed (see text).
      The 90\% confidence level for {\it ROSAT} and $2\sigma$ level for {\it XMM-Newton}.}
  \end{center}
\end{table*}

\begin{figure*}
  \begin{center}
    \includegraphics[width=10cm]{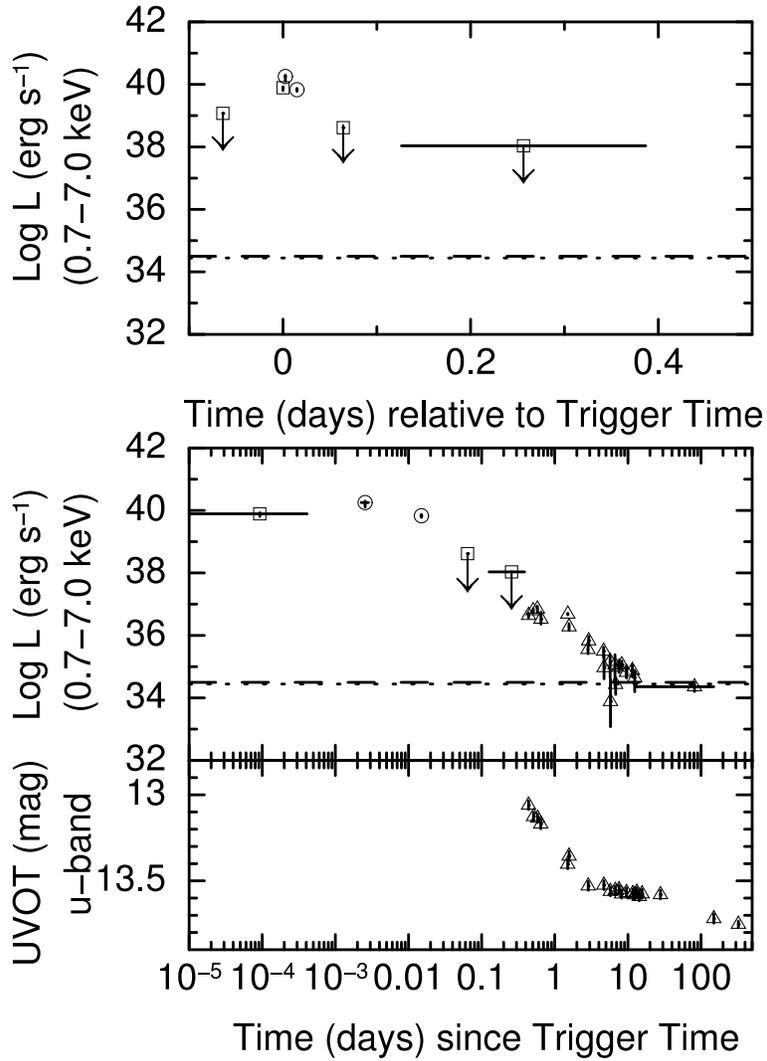}
  \end{center}
  \caption{Evolution of the fluxes of MAXI J0158$-$744.
    ({\bf Top, Middle}) The X-ray luminosity with horizontal axes
    in linear (top) and log (middle) scales.
    Here, the source distance is assumed to be the SMC distance of 60 kpc.
    ({\bf Bottom}) The $u$ band magnitude.
    The squares, circles and triangles show the data obtained by 
    MAXI GSC (Sections~\ref{sec: MAXI/GSC} and \ref{subsec: upper limit GSC}),
    MAXI SSC (Sections~\ref{sec: MAXI/SSC} and \ref{sec: MAXI/SSC spec M+2 ph})
    and {\it Swift} (Section~\ref{sec: follow-up}),
    respectively.
    For the MAXI data, the luminosities converted from the fluxes 
    of Table~\ref{table: maxi obs} and Section~\ref{subsec: upper limit GSC}
    are shown.
    For the {\it Swift} XRT data,
    those obtained by blackbody fits are shown (Section~\ref{sec: follow-up}).
    The vertical error bars represent $1\sigma$ level.
    Horizontal dashed and dotted lines show the 90\% confidence upper limits
    obtained by XMM-Newton slew survey and ROSAT PSPC, respectively,
    assuming SSS Model (Section~\ref{sec: history}).
    \label{fig: lc}
  }
\end{figure*}

\section{Discussion}\label{sec: discussion}

The X-ray transient MAXI J0158$-$744 is characterized by
(1) a soft X-ray spectrum with most of
the X-ray photons being detected below 4 keV
(Fig.~\ref{fig: maxi_finding_chart} and Fig.~\ref{fig: maxi spec}),
(2) a short duration
(between $t_2 - t_1 = 1.3 \times 10^3$~s 
and $t_3 - t_{-1} = 1.10 \times 10^4$~s; Table~\ref{table: maxi obs}),
(3) a very rapid rise time
($< t_1 - t_{\rm -1} = 5.5 \times 10^3$~s; Table~\ref{table: maxi obs}),
and (4) a huge peak luminosity of $2 \times 10^{40}$~erg~s$^{-1}$
in the $0.7-7.0$~keV band
recorded at the second MAXI scan.
The unusually soft spectrum of the outburst
is beyond astronomers' expectations, because
most short-lived luminous transient X-ray sources are hard X-ray emitters,
e.g. Gamma-ray Bursts \citep{Band+1993},
Soft Gamma Repeaters \citep{Woods_Thompson_2006},
Super-giant Fast X-ray Transients \citep{Sguera+2006},
and X-ray bursts \citep{Galloway+2008}.
The discovery of MAXI J0158$-$744, together with supernova shock
breakouts \citep{Soderberg+2008},
suggests that the wide-field monitoring experiments in soft X-rays ($<2$ keV)
will open new discovery fields.

\cite{Li+2012} reported the optical spectra obtained by SAAO and ESO,
showing that the source is a Be star. In addition, they showed that
the radial velocity of this source is consistent with the SMC,
which strongly supports that this source is located in the SMC.
Due to the similarity between the  {\it Swift} X-ray spectra and the SSS phase of novae,
they concluded that this source is a binary system consisting of a WD and a Be star.
We also analyzed the SED of the optical counterpart and found that it is consistent with that of a B-type star.
The optical spectrum taken by SMARTS showed clear emission lines of $H_\alpha$ and $H_\beta$,
confirming the conclusion of \cite{Li+2012}.

We fit the {\it Swift} XRT spectra with absorbed blackbody or Mekal models.
Neither model was strongly preferred from the statistics.
However, the evolution of the temperature of the Mekal model
shows an unexplained decrease at 3$-$7 days,
while the temperature and radius of the blackbody fits
can be understood as being due to the shrinking photosphere.
Here, the radius decreased from $\sim 10^4$~km to $\sim 100$~km,
while the temperature increased from $\sim 0.06$~keV to $\sim 0.1$~keV.
Therefore, we conclude that the spectra 
were basically blackbody-like,
and hence can be identified as a super-soft X-ray phase of a nova.
This conclusion is the same as \citet{Li+2012}.
Fits using WD atmosphere models \citep{Rauch+2010,van-Rossum_2012}
would allow further insights,
although a trial with the T{\"u}bingen WD model failed to improve 
the fits, due to the computational upper limit for the temperature \citep{Li+2012}.
Probably, more sophisticated spectral models like atmosphere models would improve the fits.
\citet{Li+2012} reported that 
adding a broad emission line at $\sim$0.7 keV and an absorption edge at $0.89$ keV
improves the fit for the spectrum at $1.54$~d significantly,
which also supports our interpretation that the early X-ray emission
is due to the SSS phase of novae.

The SSS phase spectra from the first
{\it Swift} XRT follow-up observation at $0.44$~d and
the simultaneous optical enhanced emission obtained by {\it Swift} UVOT
are unusual in the evolution of standard novae.
Nonetheless, if the optical enhanced emission is
the photospheric emission of nova ejecta as in standard novae,
the speed class indicator parameter of $t_{d, 2}$
means that it is the second fastest nova next to U Sco
\cite[$1.2$ days;][]{Schwarz+2011}
and an equal record to V838 Her \cite[$2$ days;][]{Schwarz+2011}.
The peak absolute magnitude ($-5.04 \pm 0.07$) at 0.44 day
in the first {\it Swift} UVOT observation is
four magnitudes fainter than those of typical novae ($-9.016 \pm 0.008$),
expected from the MMRD (maximum magnitude - rate of decline)
relation \citep{della-Valle_Livio_1995}.
If this enhanced optical emission was photospheric emission
from the nova ejecta,
it suggests a low ejecta mass in this nova explosion.
In the subsequent discussion below, we will show that the enhanced optical 
emission is not photospheric emission, however, the conclusion of 
the low ejecta mass remains correct.

\subsection{Shock heating mechanism}

Optically-thin thermal X-ray emission of novae is usually explained
by a shock heating mechanism at a blast wave produced by the nova explosion.
The recurrent nova, RS Ophiuchi, in 2006 exhibited
the most luminous optically-thin thermal X-ray emission 
($1 \times 10^{36}$ erg s$^{-1}$) among novae
that has ever been observed \citep{Sokoloski+2006}.
The luminosity of the X-ray outburst of MAXI J0158$-$744
was, surprisingly, four orders of magnitude larger than this.
\citet{Li+2012} explained the luminosity of MAXI J0158$-$744
by the shock heating mechanism, however, their explanation has difficulties as follows.
They tried to explain the observed luminosity of $\sim 10^{39}$ erg s$^{-1}$
at the time of the first GSC scan
(scan-ID M+0; Table~\ref{table: maxi obs})
using Equation (1) of \citet{Li+2012},
where the radius of the shock wave $r_s \sim 1.5 \times 10^{13}$~cm is assumed.
However, in order for the shock wave to expand to this radius
within $\Delta t_r = t_1 - t_{\rm -1} = 5.5 \times 10^3$~s (Table~\ref{table: maxi obs}),
the shock wave velocity must be exceptionally large
($V_s = r_s / \Delta t_r \sim 3 \times 10^{4}$ km s$^{-1}$) for novae.
In addition, if the velocity of the shock wave was such a large value,
the temperature of the plasma thermalized by the shock wave becomes very high
(
$kT = \frac{3 m_p \mu {V_s}^2}{16} \sim 1$~MeV,
where $m_p$ is the proton mass, and $\mu$ is the mean molecular weight
\footnote{
$\mu = \frac{A_{r1}}{f_{a0} + f_{a1}} = 0.61$ for a typical SMC abundance
\citep[\texttt{abund} $= 0.1$;][]{Carrera+2008}.
Here, $f_{a0} = \sum_i a_i = 1.10$, $f_{a1} = \sum_i Z_i a_i = 1.20$,
and $A_{r1} = \sum_i A_{r, i} a_i = 1.40$\citep{Anders_Grevesse_1989},
where $Z_i$, $a_i$, and $A_{r, i}$ are
the atomic number, abundance, and relative atomic mass of the $i$-th element.
The abundance is defined by the ratio of the number densities of the $i$-th element and hydrogen
($a_i = n_{{\rm X}_i} / n_{\rm H}$).
}),
which is contradictory to the observed soft spectrum 
of the outburst ($kT \sim 1$~keV).

We further discuss the shock heating scenario,
considering the very rapid rise time within $\Delta t_r$~s
and the observed low temperature ($0.97$ keV; Table~\ref{table: maxi spectrum}).
We set the onset time of the nova explosion $t_0$ between
the last scan time of the scan-ID M$-1$
($t_{\rm -1}$; Table~\ref{table: maxi obs})
and
the first scan time of the scan-ID M+0
($t_1$; Table~\ref{table: maxi obs}).
The elapsed time from $t_0$ to $t_1$ is
$t_{{\rm e}, 1} = t_1 - t_0 < \Delta t_r$~s.
We check whether the observed emission measure 
(${\it EM} \equiv \int n_e n_p dV = 1.0 \times 10^{63}\, {\it EM}_{63}\, {d_{60}}^2 $ cm$^{-3}$, 
where ${\it EM}_{63} = 2.4$ for the Mekal model fit
with free \texttt{abund} in Table~\ref{table: maxi spectrum}
and $d_{60} \simeq 1$;
here $n_e$ and $n_p$ are the number densities of electrons and protons, respectively)
can be produced at $t_{{\rm e}, 1}$
by considering two simple geometries, a filled sphere and spherical shell.
Here, we assume the circumbinary space is filled
with a fully-ionized electron-ion plasma with a constant density.
We assume the constant shock velocity $V_s$,
and then the distance ($R_1$) of the shock wave front
reaching from the surface of the WD at the time $t_{{\rm e}, 1}$ is 
\begin{equation}
R_1 = t_{{\rm e}, 1} \times V_s < \Delta t_r \times V_s = 5.5 \times 10^{11} V_{s3} \,\,\, {\rm cm},
\label{eq: R1}
\end{equation}
where $V_s = 10^3 V_{s3}$ km s$^{-1}$ ($V_{s3} \simeq 1$).

If the shape of the emission region is a filled sphere,
the emission measure at $t_{{\rm e}, 1}$ is written as
${\it EM} = \frac{4}{3} \pi {R_1}^3 {n_p}^\prime {n_e}^\prime
= \frac{4}{3} \pi {R_1}^3 f_{a 1} {{n_p}^\prime}^2$
(hereafter, we add ``~ $^\prime$ ~'' for the physical value after the shock).
Since we assume that the initial X-ray outburst is an optically-thin emission,
the condition of the optical depth is
${n_e}^\prime \sigma_T R_1 = f_{a1} {n_p}^\prime \sigma_T R_1 < 1$,
where $\sigma_T$ is the Thomson cross section.
By removing ${n_p}^\prime$,
we obtain 
$ R_1 > \frac{3 {\it EM}}{4 \pi} \sigma_T^2 f_{a1} = 1.1 \times 10^{14} f_{a1} {\it EM}_{63} {d_{60}}^2 ~{\rm cm}$.
Therefore, it is impossible to produce the observed 
emission measure at $t_1$ in the usual shock velocity
\citep[$V_{s3} \sim 1 - 10$;][]{Schwarz+2011,Warner_2008},
even at the speed of light.

Next, if the shape of the emission region is a spherical shell with a depth of $\delta R_1$,
the emission measure at $t_{{\rm e}, 1}$ is written as ${\it EM} = 4 \pi {R_1}^2 \delta R_1 {n_p}^\prime {n_e}^\prime$.
The condition of the optical depth is ${n_e}^\prime \sigma_T \delta R_1 = f_{a1} {n_p}^\prime \sigma_T \delta R_1 < 1$.
By removing ${n_p}^\prime$, 
we obtain 
$R_1 > \frac{{\it EM}}{4 \pi} \sigma_T^2 \frac{\delta R_1}{R_1} f_{a1}
 = 3.6 \times 10^{13} f_{a1} {\it EM}_{63} {d_{60}}^2 \frac{\delta R_1}{R_1} ~{\rm cm}$. 
Using Condition~\ref{eq: R1}, 
the depth of the shell is limited to
$\frac{\delta R_1}{R_1} < 1.5 \times 10^{-2} {d_{60}}^{-2} V_{s3} {{\it EM}_{63}}^{-1} f_{a1}^{-1}$.
Therefore, the emission region is a thin shell.
On the other hand, removing $\delta R_1$ and using Condition~\ref{eq: R1}, we obtain
\begin{equation}
{n_p}^\prime > \frac{ {\it EM} \sigma_T}{4 \pi R_1^2} > \frac{ {\it EM} \sigma_T}{4 \pi (\Delta t_r V_s)^2}
 = 1.8 \times 10^{14} {d_{60}}^{2} {V_{s3}}^{-2} {\it EM}_{63}  ~~{\rm cm}^{-3}.
\label{eq: n_p shell}
\end{equation}

In this high density, the shock velocity can be estimated simply 
from the observed temperature of $1.0 {\it kT}_{1.0} $ keV
(${\it kT}_{1.0} = 0.97$; Table~\ref{table: maxi spectrum})
using the shock condition 
by $V_s = \left( \frac{16 k_B T_d}{3 m_p \mu} \right)^{1/2} = 7.1 \times 10^2 {{\it kT}_{1.0}}^{1/2} \mu^{-1/2} ~ {\rm km}~{\rm s}^{-1}$,
where $T_d$ is the temperature of the shock-heated gas in the down-stream region.

The radiative cooling time scale by free-free process is 
\begin{equation}
t_{\rm cool} = 3 \left( \frac{3 m_e}{2 \pi} \right)^{1/2} \frac{3 h m_e c^3}{2^5 \pi e^6}
(k_B T_e)^{1/2} {n_p^\prime}^{-1} \bar{g_B}^{-1} f_{b}
= 2.1 \times 10^{19} (k_B T_e)^{1/2} {n_p^\prime}^{-1} f_{b},
\label{eq: tcool}
\end{equation}
where $h$, $c$, and $\bar{g_B}$ are
Planck constant, speed of light, and averaged Gaunt factor, respectively,
and we set $\bar{g_B} = 1.2$ \citep{Rybicki_Lightman_1979}.
Here, $f_{b} = \frac{f_{a0} + f_{a1}}{2 f_{a1} f_{a2}}$ ($f_{a2} = \sum_i Z_i^2 a_i$) is 0.68
for a typical SMC abundance.
From Condition~\ref{eq: n_p shell}, the $t_{\rm cool}$ is limited to
$t_{\rm cool} < 4.6 ~{d_{60}}^{-2} V_{s3}^2 {{\it EM}_{63}}^{-1} {{\it kT}_{1.0}}^{1/2} f_b$~s.
Since the cooling time scale is short, the width of the emitting shell is simply written as
$ \delta R_1 = \frac{1}{4} V_s \times t_{\rm cool}$,
where $\frac{1}{4} V_s$ is the velocity of the post-shock region in the rest frame of the shock wave.
Using $\delta R_1 = \frac{{\it EM}}{4 \pi R_1^2 {n_p^\prime}^2 f_{a1}}$ and Equation~\ref{eq: tcool},
we derive the relation between $n_p^{\prime}$ and $R_1$,
$n_p^\prime = 3.8 \times 10^{39} \times R_1^{-2} {d_{60}}^2 {V_{s3}}^{-1} {\it EM}_{63} {{\it kT}_{1.0}}^{-1/2}
f_{a1}^{-1} f_{b}^{-1} ~~{\rm cm}^{-3}.$
Using Condition~\ref{eq: R1}, the density is limited to
$n_p^\prime > 1.3 \times 10^{16} {d_{60}}^2 {V_{s3}}^{-3} {\it EM}_{63} {{\it kT}_{1.0}}^{-1/2} 
f_{a1}^{-1} f_{b}^{-1} ~~{\rm cm}^{-3}.$
By setting ${\it kT}_{1.0} = 0.97$, ${\it EM}_{63} = 2.4$, $d_{60} = 1$ (Table~\ref{table: maxi spectrum})
and assuming a typical SMC abundance (\texttt{abund} $= 0.1$),
we obtain
$n_p^\prime > 5.2 \times 10^{16} ~~{\rm cm}^{-3}$.
This density is much larger than that in the stellar wind
and even that in a circumstellar equatorial disk
around a Be star, typically $\lesssim 10^{13}$ cm$^{-3}$ \citep{Waters+1988}.
Therefore, we conclude that the shock heating scenario cannot explain the soft X-ray outburst observed by MAXI.

\subsection{Thermonuclear runaways at the ignition phase}

We thus instead propose another scenario
to explain this outburst by invoking an extraordinary massive WD.
A more massive WD has a smaller radius \citep{Nauenberg+1972},
and thus a higher surface gravity leading to a higher pressure in the accumulated mass.
The nova explosion on a massive WD is triggered by less fuel,
and thus it results in a short nova duration.
The observed SSS phase of MAXI J0158$-$744
started much earlier ($< 0.44$ days) and lasted a much shorter time ($\sim$ one month)
than other fast novae \citep{Hachisu_Kato_2006, Schwarz+2011, Li+2012}.
The earliest turn-on of a SSS phase observed so far was 10 days
in U Sco \citep{Schwarz+2011}
and $11 \pm 5$ days in one of 60 novae in M31 \citep{Henze+2011}.
The extremely early SSS phase of MAXI J0158$-$744
is unexpected in models of novae on typical solar mass WDs \citep{Hachisu_Kato_2006}.
It suggests an unusually low ejecta mass in the nova explosion,
and thus unusually massive white dwarf near the Chandrasekhar mass.
It might even suggest a super-Chandrasekhar mass.
Indeed, according to theoretical models \citep{Yoon_Langer_2004, Hachisu+2012},
WDs can acquire super-Chandrasekhar masses up to $2.3-2.7M_\odot$,
if they rotate differentially.

\begin{figure*}
  \begin{center}
    \includegraphics[width=15cm]{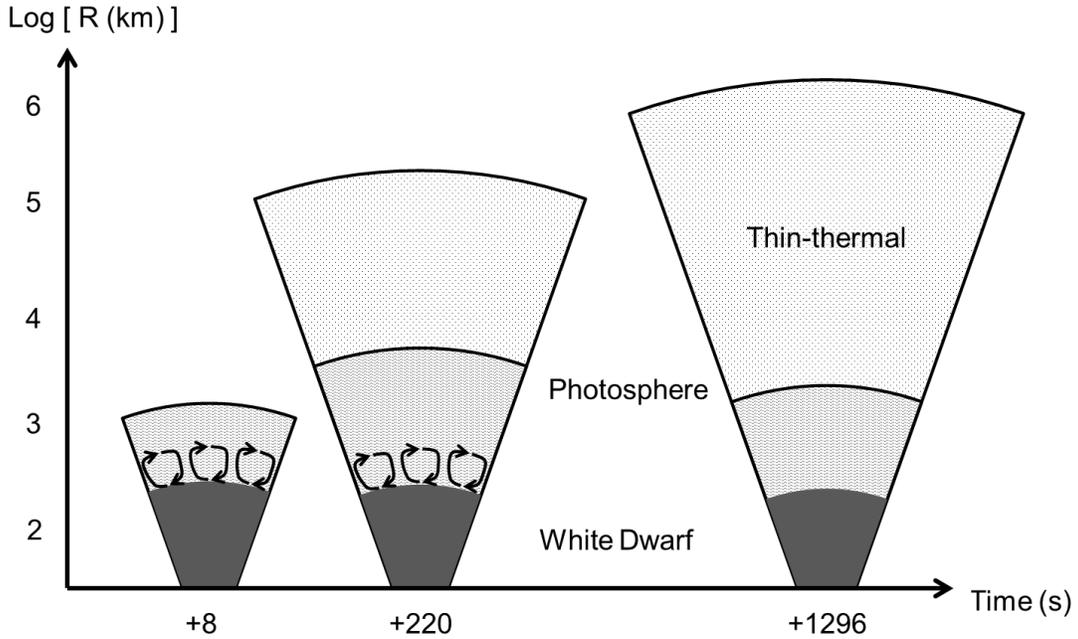}
  \end{center}
  \caption{Schematic view of the ignition phase of the nova, MAXI J0158$-$744.
    The super-Eddington luminosity is
    produced by convection during the first $\sim 100$~s.
    The neon emission line at $+1296$~s was produced from the
    optically-thin thermal region filled with the ejecta dredged-up from the O-Ne WD.
    \label{fig: nova evolution}
    }
\end{figure*}

With this new perspective, 
we propose to interpret the initial super-Eddington X-ray outburst as 
an ignition phase of a nova just after the TNR, a fireball phase
\citep{Starrfield_Iliadis_Hix_2008,Starrfield+1998,Krautter_2008a,Krautter_2008b}.
In this process, the thermal energy produced by the TNR is conveyed by the convection
and released outside the envelope with a timescale of $\sim 100$~s,
characterized by the half-lives of unstable nuclei (Fig.~\ref{fig: nova evolution}).
In novae on a white dwarf with a usual mass,
transient soft X-ray emission ($<~ 0.1$~keV)
for $\sim 100 - 1000$~s just after the TNR
is theoretically expected \citep{Starrfield_Iliadis_Hix_2008},
but that has not been observed yet.
It is expected to reach about 10 times the Eddington luminosity
\citep{Starrfield_Iliadis_Hix_2008}.
For a very massive WD, we speculate that the TNR would produce
more luminous X-ray emission with higher temperature
due to a smaller amount of the envelope at the ignition phase of a nova.

In this scenario, blackbody-like X-ray emission is expected at the ignition phase.
In the spectral analysis (Table~\ref{table: maxi spectrum}),
we obtained the radius of the photosphere to be $r_{\rm ph} \sim 10^3 r_{\rm ph, 8}$~km
($r_{\rm ph, 8} \simeq 1$).
The rate of mass ejection ($\dot{M}$) can be estimated from this radius as follows.
From the continuity equation for the distribution of ejecta around the WD,
$4 \pi r^2 \rho V_e = \dot{M}$,
where $\dot{M}$ is a rate of mass ejection from the WD and constant in the radial distance ($r$),
and $\rho = \sum A_{r, i} a_i m_p n_p = A_{r 1} m_p n_p$ is mass density,
the number density of protons is written by
\begin{equation}
  n_p = \frac{\dot{M}}{4 \pi r^2 A_{r 1} m_p V_e}.
  \label{eq: n_p}
\end{equation}
The optical depth condition is written by
\begin{equation}
  \tau = \int_{r_{\rm ph}}^\infty n_e \sigma_T dr 
  = \int_{r_{\rm ph}}^\infty f_{a 1} n_p \sigma_T dr 
  = \frac{f_{a 1}}{A_{r 1}} \frac{\dot{M} \sigma_T}{4 \pi m_p V_e r_{\rm ph}} = 1.
\end{equation}
Here, $f_{a 1} / A_{r 1} \simeq Z_{10} / A_{r, 10} \simeq 0.5$
for large \texttt{abund} of neon.
Therefore, $\dot{M}$ is obtained to be
\begin{equation}
  \dot{M} = \frac{A_{r 1}}{f_{a 1}} \frac{4 \pi m_p V_e r_{\rm ph}}{\sigma_T} 
  \simeq 6.4 \times 10^{17} V_{e3} r_{\rm ph, 8} \,\,\, {\rm g}\,\,{\rm s}^{-1}.
  \label{eq: m_dot}
\end{equation}

On the other hand, the reaction rate of mass producing nuclear energy ($\dot{M_f}$)
is related to the observed luminosity ($L_X$) by $\eta \dot{M_f} c^2 > L_X$,
where $\eta = 0.007$ and $L_X = 10^{40}$ erg s$^{-1}$ (Table~\ref{table: maxi spectrum}).
Then, $\dot{M_f} > L_X / \eta c^2 = 2 \times 10^{21}$ g s$^{-1}$.
Therefore, the relation $\dot{M_f} >> \dot{M}$ is obtained, which
means that energy produced by the TNR
at the bottom of accreted layer can escape as an X-ray photons efficiently
with very small mass ejection, despite the super-Eddington luminosity.

Thus, there must be some sort of mechanism
to realize the super-Eddington luminosity with a small mass ejection.
We infer a convection just after the TNR
\citep{Starrfield_Iliadis_Hix_2008,Starrfield+1998}
for that mechanism, then we expect that future theoretical works
of the TNR process,
applied to the mass range near or over the Chandrasekhar limit,
will clarify this mechanism.
We also suspect that photon bubbles
in highly magnetized atmospheres \citep{Begelman_2001} 
may work to solve this problem.
According to \cite{Begelman_2001},
to produce the $\sim 100$ Eddington luminosity with small mass ejection,
the magnetic pressure $P_{\rm mag}$ must be $\sim 100$ times larger than
the gas pressure $P_{\rm gas}$. On the other hand, 
the gas pressure at the bottom of an accreted gas layer at an ignition of a nova
is expected to be $P_{\rm gas} \sim 10^{20}$ dyne cm$^{-2}$
\citep{Starrfield_Iliadis_Hix_2008,Fujimoto_1982}.
Then the magnetic field ($B$) necessary for the $\sim 100$ Eddington luminosity is
$B = \left( 8 \pi P_{\rm mag} \right)^{1/2} \sim \left( 8 \pi 100 P_{\rm gas} \right)^{1/2}
\sim 5 \times 10^{11}$ G.
Interestingly, such highly magnetized WDs with super-Chandrasekhar masses
($2.3-2.6 M_\odot$)
are predicted theoretically \citep{Das_Mukhopadhyay_2012}.

Since the TNR process is expected to last for $\sim 100$~s
at the bottom of the accreted mass layer
on the surface of WDs \citep{Starrfield_Iliadis_Hix_2008,Starrfield+1998},
the rate of mass ejection probably peaked 
between the scans at $+220$~s and $+1296$~s.
It means that MAXI scans at $+8$~s and $+220$~s observed 
the photospheric expansion phase
\citep[B $\rightarrow$ C in Fig.~1 of][]{Kato_Hachisu_1994},
while the MAXI scan at $+1296$~s observed 
the shrinking phase (C $\rightarrow$ D in the same figure).
The strong neon emission line at $+1296$~s suggests that
there was an optically-thin thermal emission region surrounding the photosphere
and filled with ejecta dredged-up from a massive O-Ne WD.
Such ejecta may have been provided
by the previous photospheric expansion.
It must be noted that the existing models of the TNR do not predict
this surrounding emission line region.
This observation provides us new physical details.

In this scenario, the optical enhancement observed by {\it Swift} UVOT
is no longer the usual photospheric emission of nova ejecta.
Since the optical decay seems correlated with the decay of SSS X-ray emission
(Fig.~\ref{fig: lc}, middle and bottom),
it can be explained by the reprocessed emission from the X-ray irradiated
circumstellar disk of the Be star.
It is justified by the fact that
the size of the optical enhanced emission
($6.6 \times 10^{11}$~cm; Section~\ref{sec: follow-up}) is
comparable to the disk scale height \citep{Zorec+2007}.

\section{Summary}\label{sec: summary}
Monitor of All-sky X-ray Image (MAXI) discovered
an extraordinarily luminous soft X-ray transient,
MAXI J0158$-$744, near the Small Magellanic Cloud (SMC)
on 2011 November 11. This source is a binary system consisting
of a white dwarf (WD) and a Be star
at the distance of the SMC. 
MAXI detected it in three scans at $+8$~s, $+220$~s and $+1296$~s
after the trigger time.
The X-ray luminosity peaked on the second scan
at $2 \times 10^{40}$ erg s$^{-1}$ ($0.7-7.0$ keV),
which is two orders of magnitude
brighter than the Eddington luminosity of a solar mass object.
The spectrum of the third scan showed a He-like neon emission, suggesting
that the emission contains an optically-thin thermal component
and the WD is a massive O-Ne WD.
While the X-ray outburst could be considered
as a kind of nova on the basis of the luminosity and the spectral evolutions,
the huge peak luminosity and the rapid rise time ($< 5.5 \times 10^3$~s)
are difficult to explain by shock-induced emission,
accepted for optically-thin thermal emission in nova explosions
observed so far.
Instead, we propose the scenario that the X-ray outburst
is the direct manifestation of the 
thermonuclear runaway process
at the onset of the nova explosion,
the so-called fireball phase.
The super-Eddington X-ray outburst
and the subsequent very early 
super-soft source phase
indicate a small ejecta mass,
implying the underlying WD is unusually massive
near the Chandrasekhar limit, or possibly exceeding the limit.

\acknowledgments

We are grateful to the members of the MAXI and {\it Swift} operation teams.
We thank K.~Asano, I.~Hachisu, D.~N.~Burrows, D.~Takei, S.~R.~Kulkarni, Y.~Maeda,
and T.~Shigeyama for discussions and comments.
This research was partially supported by the Ministry of Education, Culture, Sports, Science
and Technology (MEXT), Grant-in-Aid No.
23740147, 19047001,
and Global-COE from MEXT ``The Next Generation of Physics, Spun from Universality and Emergence''
and ``Nanoscience and Quantum Physics.''
JPO, KLP and NPMK acknowledge financial support from the UK Space Agency.
JAK acknowledges support from NASA.
This work was supported by the Australian Research Council
Discovery Projects funding scheme (project number DP120102393).

\end{document}